\documentclass[aip,jmp,amsmath,amssymb,preprint,author-numerical]{revtex4-1}
\usepackage{amsthm}
\usepackage{graphicx}
\usepackage{dcolumn}
\usepackage{bm}
\usepackage[mathlines]{lineno}

\usepackage[utf8]{inputenc}
\usepackage[T1]{fontenc}
\usepackage{mathptmx}
\usepackage{hyperref}

\newtheorem{thm}{Theorem}

\newtheorem{conj}{Conjecture}

\newtheorem{cor}{Corollary}

\newtheorem{lem}{Lemma}

\theoremstyle{definition}

\theoremstyle{remark}
\newtheorem{rem}{Remark}

\begin{document}


\title{The squashed entanglement of the noiseless quantum Gaussian attenuator and amplifier}

\author{Giacomo De Palma}
\email{giacomo.depalma@math.ku.dk}
\affiliation{QMATH, Department of Mathematical Sciences, University of Copenhagen, Universitetsparken 5, 2100 Copenhagen, Denmark}%

\date{\today}

\begin{abstract}
We determine the maximum squashed entanglement achievable between sender and receiver of the noiseless quantum Gaussian attenuators and amplifiers, and prove that it is achieved sending half of an infinitely squeezed two-mode vacuum state.
The key ingredient of the proof is a lower bound to the squashed entanglement of the quantum Gaussian states obtained applying a two-mode squeezing operation to a quantum thermal Gaussian state tensored with the vacuum state.
This is the first lower bound to the squashed entanglement of a quantum Gaussian state, and opens the way to determine the squashed entanglement of all quantum Gaussian channels.
Moreover, we determine the classical squashed entanglement of the quantum Gaussian states above, and show that it is strictly larger than their squashed entanglement.
This is the first time that the classical squashed entanglement of a mixed quantum Gaussian state is determined.
\end{abstract}

\maketitle

\section{Introduction}
The squashed entanglement of a bipartite quantum state $\rho_{AB}$ is the infimum over all its possible extensions $\rho_{ABR}$ of half of the quantum mutual information between the quantum systems $A$ and $B$ conditioned on the quantum system $R$ \cite{tucci1999quantum,tucci2000separability,tucci2000entanglement,tucci2001relaxation,tucci2001entanglement,tucci2002entanglement,christandl2004squashed,brandao2011faithful,seshadreesan2015renyi}:
\begin{equation}\label{eq:defEsq}
E_{\mathrm{sq}}(\rho_{AB}) = \tfrac{1}{2}\inf\left\{I(A;B|R)_{\rho_{ABR}}:\mathrm{Tr}_R\rho_{ABR}=\rho_{AB}\right\}\,.
\end{equation}
Here the quantum conditional mutual information is defined as \cite{wilde2017quantum}
\begin{equation}
I(A;B|R) = S(A|R) + S(B|R) - S(AB|R)\,,
\end{equation}
where $S(X|Y)$ is the quantum conditional entropy \cite{wilde2017quantum,holevo2013quantum,tomamichel2015quantum}.

The squashed entanglement is one of the two main entanglement measures in quantum communication theory: together with the relative entropy of entanglement \cite{audenaert2001asymptotic,vedral2002role}, it provides the best known upper bound to the length of a shared secret key that can be generated by two parties holding many copies of the quantum state \cite{christandl2004squashed,christandl2007unifying,li2014relative,wilde2016squashed}.
Moreover, it has applications in recoverability theory \cite{seshadreesan2015fidelity,li2018squashed} and multiparty information theory \cite{adesso2007coexistence,avis2008distributed,yang2009squashed}.

Any entanglement measure for quantum states can be extended to quantum channels defining it as the maximum entanglement achievable between sender and receiver.
The relative entropy of entanglement of several quantum channels has been determined in \onlinecite{pirandola2017fundamental}.
The squashed entanglement of a quantum channel $\Phi$ \cite{takeoka2014squashed} is the maximum squashed entanglement achievable between sender and receiver:
\begin{equation}
E_{\mathrm{sq}}(\Phi) = \sup_{\rho_{AB}}E_{\mathrm{sq}}\left((\mathbb{I}_A\otimes\Phi)(\rho_{AB})\right)\,,
\end{equation}
where the sender generates the bipartite quantum state $\rho_{AB}$, keeps the quantum system $A$ and sends the quantum system $B$ to the receiver through $\Phi$.
In the same way as the squashed entanglement of a quantum state is an upper bound to the distillable key of the state, the squashed entanglement of a quantum channel is an upper bound to the capacity of the channel to generate a secret key shared between sender and receiver \cite{berta2018amortization,takeoka2014squashed}.

We prove a lower bound on the squashed entanglement of the quantum Gaussian states obtained applying a two-mode squeezing operation to a thermal quantum Gaussian state tensored with the vacuum state (\autoref{thm:squashedrho}; see \eqref{eq:defrho} for the definition of the states).
This is the first lower bound to the squashed entanglement of a quantum Gaussian state.
Previous results restrict the optimization in \eqref{eq:defEsq} to Gaussian extensions and consider the 2-R\'enyi entropy instead of the von Neumann entropy \cite{lami2017log}.
Our bound is optimal in the limit of infinite energy and extremely tight already from one average photon (\autoref{fig:bounds}).
Lower bounds to the squashed entanglement are notoriously difficult to prove, since the optimization in \eqref{eq:defEsq} over all the possible extensions of the quantum state is almost never analytically treatable.
We overcome this difficulty with the quantum conditional Entropy Power Inequality \cite{koenig2015conditional,de2018conditional,huber2018conditional,de2019entropy}, which holds for any conditioning quantum system.

We apply \autoref{thm:squashedrho} to prove our main result: a new lower bound to the squashed entanglement of the noiseless quantum Gaussian attenuators and amplifiers (\autoref{thm:main}).
This lower bound matches the upper bound proven in \onlinecite{takeoka2014fundamental,takeoka2014squashed,goodenough2016assessing,davis2018energy}.
Therefore, \autoref{thm:main} determines the maximum squashed entanglement achievable between sender and receiver for the noiseless quantum Gaussian attenuators and amplifiers, and proves that it is achieved sending half of an infinitely squeezed two-mode vacuum state.
The maximum achievable squashed entanglement is proved to be $\ln\frac{1+\eta}{1-\eta}$ for the attenuator with attenuation parameter $0\le\eta\le1$ and $\ln\frac{\kappa+1}{\kappa-1}$ for the amplifier with amplification parameter $\kappa\ge1$.
The noiseless quantum Gaussian attenuator and amplifier play a key role in quantum communication theory.
The amplifier provides the mathematical model for the amplification of electromagnetic signals, and the attenuator provides the mathematical model for the propagation of electromagnetic signals through optical fibers \cite{caruso2006one,caruso2006degradability,holevo2007one,weedbrook2012gaussian,serafini2017quantum}, which are the main platform for quantum key distribution and for the transmission of quantum states in the forthcoming quantum internet \cite{bennett1984proceedings,ekert1991quantum,gisin2002quantum,lloyd2004infrastructure,braunstein2005quantum,kimble2008quantum, scarani2009security,weedbrook2012gaussian,pirandola2016physics,pirandola2016capacities,azuma2016fundamental,laurenza2017general, laurenza2018finite,cope2018converse,wehner2018quantum,laurenza2019tight,pirandola2019end,pirandola2019advances}.
A proof based on the relative entropy of entanglement and on the teleportation stretching technique has determined the capacity of the noiseless quantum Gaussian attenuator and amplifier to generate a shared secret to be $\ln\frac{1}{1-\eta}$ and $\ln\frac{\kappa}{\kappa-1}$ nats per channel use, respectively \cite{pirandola2009direct,pirandola2017fundamental} (later a generalization to repeaters \cite{pirandola2016capacities,pirandola2019end} was proved, and strong converse bounds \cite{wilde2017converse,pirandola2018theory,christandl2017relative} were also considered).
\autoref{thm:main} proves that these channels can generate a squashed entanglement between sender and receiver strictly larger than their secret key capacity.

A fundamental question about the squashed entanglement of a quantum state is whether the minimization in \eqref{eq:defEsq} can be restricted to classical extensions \cite{tucci2002entanglement}.
The answer is known to be negative in general \cite{brandao2008entanglement}, and this has led to the definition to the classical squashed entanglement \cite{yang2007conditional,yang2008additive,yang2009squashed,song2009lower,li2014relative,huang2014computing}, which has later found an operational interpretation as the minimum cost of classical communication required for assisted entanglement dilution \cite{wakakuwa2019communication}.
We determine the classical squashed entanglement of the quantum Gaussian states \eqref{eq:defrho} and prove that it is achieved by a Gaussian extension and strictly larger than their squashed entanglement, with equality only when the state is pure or separable (\autoref{thm:Esqc}, \autoref{fig:bounds}).
This is the first time that the classical squashed entanglement of a mixed quantum Gaussian state is determined.
The proof is based on the one-mode version of the constrained minimum output entropy conjecture for the noiseless quantum Gaussian amplifier and for its complementary channel \cite{guha2007classical,guha2007classicalproc,guha2008entropy,guha2008capacity,konig2014entropy,konig2016corrections,de2014generalization, de2015multimode,de2015passive,de2016passive,de2016gaussian,de2016gaussiannew,de2017gaussian,de2017multimode,qi2017minimum,de2018pq,de2018gaussian,de2019new}.
\autoref{thm:Esqc} also proves that the multi-mode version of the conjecture implies that the classical squashed entanglement of the quantum Gaussian states \eqref{eq:defrho} does not decrease regularizing over many copies of the state.
Therefore, assuming the multi-mode conjecture, the asymptotic classical squashed entanglement of the states \eqref{eq:defrho} coincides with their classical squashed entanglement and is strictly larger than their squashed entanglement.

The paper is structured as follows.
In \autoref{sec:QGC} we introduce quantum Gaussian systems, states and channels.
In \autoref{sec:Esq} we prove the lower bound to the squashed entanglement of the quantum Gaussian states \eqref{eq:defrho} (\autoref{thm:squashedrho}) and we determine the squashed entanglement of the noiseless quantum Gaussian attenuator and amplifier (\autoref{thm:main}).
In \autoref{sec:Esqc} we determine the classical squashed entanglement of the quantum Gaussian state \eqref{eq:defrho}.
Conclusions and open problems are presented in \autoref{sec:concl}.
In \autoref{app}, we present the entropic inequalities employed in the proofs.

\section{Quantum Gaussian systems}\label{sec:QGC}
A one-mode quantum Gaussian system is the mathematical model for a harmonic oscillator or for a mode of the electromagnetic radiation.
The Hilbert space of a one-mode quantum Gaussian system is the irreducible representation of the canonical commutation relation \cite{weedbrook2012gaussian,serafini2017quantum}, [\onlinecite[Chapter 12]{holevo2013quantum}]
\begin{equation}
\left[Q,\,P\right]=\mathbb{I}\,,
\end{equation}
where $Q$ and $P$ are the quadrature operators, which for the harmonic oscillator represent position and momentum.
The Hamiltonian that counts the number of excitations or photons is
\begin{equation}\label{eq:defH}
H=a^\dag a\,,
\end{equation}
where
\begin{equation}
a=\frac{Q+\mathrm{i}\,P}{\sqrt{2}}
\end{equation}
is the ladder operator.
The vector annihilated by $a$ is the vacuum and is denoted by $|0\rangle$.
A quantum Gaussian state is a quantum state proportional to the exponential of a quadratic polynomial in $Q$ and $P$.
The most important quantum Gaussian states are the thermal Gaussian states, where the polynomial is proportional to the Hamiltonian \eqref{eq:defH}:
\begin{equation}\label{eq:omega}
\omega(E)= \frac{1}{\left(E+1\right)}\,\left(\frac{E}{E+1}\right)^{a^\dag a}\,,
\end{equation}
and $E\ge0$ is the average energy:
\begin{equation}
\mathrm{Tr}\left[\omega(E)\,a^\dag a\right] = E\,.
\end{equation}
We notice that $\omega(0)=|0\rangle\langle0|$ is the vacuum state.
The von Neumann entropy of $\omega(E)$ is
\begin{equation}\label{eq:defg}
S(\omega(E)) = \left(E+1\right)\ln\left(E+1\right)- E\ln E =: g(E)\,.
\end{equation}
An $n$-mode Gaussian quantum system is the union of $n$ one-mode Gaussian quantum systems, and its Hilbert space is the $n$-th tensor power of the Hilbert space of a one-mode Gaussian quantum system.
Let $R_1=Q_1,\,R_2=P_1,\,\ldots,\,R_{2n-1}=Q_n,\,R_{2n}=P_n$ be the quadrature operators of the $n$ modes, satisfying the canonical commutation relations
\begin{equation}
\left[R_i,\,R_j\right] = \mathrm{i}\,\Delta_{ij}\,\mathbb{I}\,,\qquad i,\,j=1,\,\ldots,\,2n\,,
\end{equation}
where
\begin{equation}
\Delta = \bigoplus_{k=1}^n\left(
                            \begin{array}{cc}
                              0 & 1 \\
                              -1 & 0 \\
                            \end{array}
                          \right)
\end{equation}
is the symplectic form.
The covariance matrix of a quantum state $\rho$ is the $2n\times2n$ positive real matrix given by
\begin{equation}
\sigma(\rho)_{ij} = \tfrac{1}{2}\mathrm{Tr}\left[\rho\left\{R_i-\mathrm{Tr}\left[\rho\,R_i\right],\;R_j-\mathrm{Tr}\left[\rho\,R_j\right]\right\}\right]\,,\quad i,\,j=1,\,\ldots,\,2n\,,
\end{equation}
where
\begin{equation}
\{X,\,Y\} = X\,Y + Y\,X
\end{equation}
is the anti-commutator.
The eigenvalues of the matrix $\Delta^{-1}\sigma$ are pure imaginary and pairwise opposite.
Their absolute values are the symplectic eigenvalues of $\sigma$ \cite{holevo2013quantum}.
An $n$-mode quantum Gaussian state is a state proportional to the exponential of a quadratic polynomial in the quadratures.
Its von Neumann entropy is \cite{bombelli1986quantum}
\begin{equation}
S = \sum_{k=1}^n g\left(\nu_k-\tfrac{1}{2}\right)\,,
\end{equation}
where $\nu_1,\,\ldots,\,\nu_n$ are the symplectic eigenvalues of its covariance matrix, and $g$ is defined in \eqref{eq:defg}.

Quantum Gaussian channels are the quantum channels that preserve the set of quantum Gaussian states.
The most important families of quantum Gaussian channels are the beam-splitter, the squeezing and the quantum Gaussian attenuators and amplifiers.
The beam-splitter and the squeezing are the quantum counterparts of the classical linear mixing of random variables, and are the main transformations in quantum optics.
Let $A$ and $B$ be one-mode quantum Gaussian systems with ladder operators $a$ and $b$, respectively.
The \emph{beam-splitter} of transmissivity $0\le\eta\le1$ is implemented by the unitary operator
\begin{equation}\label{eq:defU}
U_\eta=\exp\left(\left(a^\dag b-b^\dag a\right)\arccos\sqrt{\eta}\right)\,,
\end{equation}
and performs a linear rotation of the ladder operators [\onlinecite[Section 1.4.2]{ferraro2005gaussian}]:
\begin{align}\label{eq:defUlambda}
U_\eta^\dag\,a\,U_\eta &= \sqrt{\eta}\,a+\sqrt{1-\eta}\,b\,,\nonumber\\
U_\eta^\dag\,b\,U_\eta &= -\sqrt{1-\eta}\,a+\sqrt{\eta}\,b\,.
\end{align}
The \emph{squeezing} \cite{barnett2002methods} of parameter $\kappa\ge1$ is implemented by the unitary operator
\begin{equation}\label{eq:defUk}
U_\kappa=\exp\left(\left(a^\dag b^\dag-a\,b\right)\mathrm{arccosh}\sqrt{\kappa}\right)\,,
\end{equation}
and acts on the ladder operators as
\begin{align}
U_\kappa^\dag\,a\,U_\kappa &= \sqrt{\kappa}\,a+\sqrt{\kappa-1}\,b^\dag\,,\nonumber\\
U_\kappa^\dag\,b\,U_\kappa &= \sqrt{\kappa-1}\,a^\dag+\sqrt{\kappa}\,b\,.
\end{align}

The noiseless quantum Gaussian attenuators model the attenuation affecting electromagnetic signals traveling through optical fibers or free space.
The one-mode \emph{noiseless quantum Gaussian attenuator} $\mathcal{E}_{\eta}$ [\onlinecite[case (C) with $k=\sqrt{\eta}$ and $N=0$]{holevo2007one}] can be implemented mixing the input state $\rho$ with the one-mode vacuum state through a beam-splitter of transmissivity $0\le\eta\le1$:
\begin{equation}\label{eq:defE}
\mathcal{E}_{\eta}(\rho) = \mathrm{Tr}_B\left[U_\eta\left(\rho\otimes|0\rangle\langle0|\right)U_\eta^\dag\right]\,.
\end{equation}
The noiseless quantum Gaussian amplifiers model the amplification of electromagnetic signals.
The one-mode \emph{noiseless quantum Gaussian amplifier} $\mathcal{A}_{\kappa}$ [\onlinecite[case (C) with $k=\sqrt{\kappa}$ and $N=0$]{holevo2007one}] can be implemented performing a squeezing of parameter $\kappa\ge1$ on the input state $\rho$ and the one-mode vacuum state:
\begin{equation}\label{eq:defA}
\mathcal{A}_{\kappa}(\rho) = \mathrm{Tr}_B\left[U_\kappa\left(\rho\otimes|0\rangle\langle0|\right)U_\kappa^\dag\right]\,.
\end{equation}

\section{Squashed entanglement}\label{sec:Esq}
Let $A$ and $B$ be one-mode quantum Gaussian systems.
For any $\kappa\ge1$ and any $E\ge0$ we consider the quantum Gaussian state
\begin{equation}\label{eq:defrho}
\rho_{AB}^{\kappa,E} = U_{\kappa}\left(\omega_A(E)\otimes|0\rangle_B\langle0|\right)U_{\kappa}^\dag\,,
\end{equation}
where $\omega_A(E)$ is the thermal quantum Gaussian state on $A$ with average energy $E$ defined in \eqref{eq:omega}, $|0\rangle_B$ is the vacuum state of $B$ and $U_{\kappa}$ is the two-mode squeezing operator on $AB$ with squeezing parameter $\kappa$ defined in \eqref{eq:defUk}.

\begin{thm}\label{thm:squashedrho}
For any $\kappa\ge1$ and any $E\ge0$, the squashed entanglement of the quantum Gaussian state $\rho_{AB}^{\kappa,E}$ defined in \eqref{eq:defrho} satisfies
\begin{equation}\label{eq:Esqt}
\ln\left(2\kappa-1\right) \le E_{\mathrm{sq}}\left(\rho_{AB}^{\kappa,E}\right) \le g\left(\left(\kappa-\tfrac{1}{2}\right)E+\kappa-1\right)-g\left(\tfrac{E}{2}\right)\,.
\end{equation}
Moreover, the gap between the upper and lower bound of \eqref{eq:Esqt} is at most $\ln\frac{\mathrm{e}}{2}\simeq0.31$, and tends to zero in the limit $E\to\infty$.
We conjecture that the upper bound of \eqref{eq:Esqt} is the actual value of the squashed entanglement of $\rho_{AB}^{\kappa,E}$.
\begin{proof}
Let $\rho_{ABR}$ be an extension of $\rho_{AB}^{\kappa,E}$.
The quantum state $U_{\kappa}^\dag\,\rho_{ABR}\,U_{\kappa}$ is an extension of $\omega_A(E)\otimes|0\rangle_B\langle0|$, therefore it has the form $\omega_{AR}\otimes|0\rangle_B\langle0|$ for some quantum state $\omega_{AR}$ that is an extension of $\omega_A(E)$.
The quantum state $\rho_{ABR}$ has then the form
\begin{equation}\label{eq:rhoABR}
\rho_{ABR} = U_{\kappa}\left(\omega_{AR}\otimes|0\rangle_B\langle0|\right)U_{\kappa}^\dag\,,\qquad\mathrm{Tr}_R\omega_{AR}=\omega_A(E)\,.
\end{equation}
Conversely, any state of the form \eqref{eq:rhoABR} is an extension of $\rho_{AB}^{\kappa,E}$.
\paragraph{Lower bound}
The quantum conditional Entropy Power Inequality (\autoref{thm:cEPI}) implies
\begin{align}\label{eq:ineqI}
I(A;B|R)_{\rho_{ABR}} &= S(A|R)_{\rho_{ABR}} + S(B|R)_{\rho_{ABR}} - S(AB|R)_{\rho_{ABR}}\nonumber\\
&= S(A|R)_{\rho_{ABR}} + S(B|R)_{\rho_{ABR}} - S(A|R)_{\omega_{AR}}\nonumber\\
&\ge \ln\left(2\kappa\left(\kappa-1\right)\cosh S(A|R)_{\omega_{AR}} + \kappa^2 + \left(\kappa-1\right)^2\right)\nonumber\\
&\ge 2\ln\left(2\kappa-1\right)\,,
\end{align}
where the last inequality is saturated when $S(A|R)_{\omega_{AR}}=0$.
The lower bound in \eqref{eq:Esqt} follows taking the infimum of the left-hand side of \eqref{eq:ineqI} over the extensions $\rho_{ABR}$.

\paragraph{Upper bound}
We consider the one-parameter family of extensions $\{\rho_{ABR}(\eta)\}_{0\le\eta\le1}$ of $\rho_{AB}^{\kappa,E}$ of the form \eqref{eq:rhoABR} where $R$ is a one-mode quantum Gaussian system and
\begin{equation}
\omega_{AR}(\eta) = (\mathbb{I}_A\otimes\mathcal{E}_\eta)(|\phi_E\rangle_{AR}\langle\phi_E|)
\end{equation}
is the quantum Gaussian state obtained applying the noiseless quantum Gaussian attenuator with attenuation parameter $\eta$ to half of the two-mode squeezed vacuum state $|\phi_E\rangle_{AR}$ with average energy per mode $E$.
We leave $\eta$ as a free parameter over which we will optimize in the end.
The covariance matrix of $\omega_{AR}(\eta)$ is
\begin{equation}\label{eq:sigmaAReta}
\sigma(\omega_{AR}(\eta)) = \left(
                \begin{array}{cc}
                  \left(E+\tfrac{1}{2}\right)I_2 & \sqrt{\eta\,E\left(E+1\right)}\,\sigma_Z \\
                  \sqrt{\eta\,E\left(E+1\right)}\,\sigma_Z & \left(\eta\,E + \tfrac{1}{2}\right)I_2 \\
                \end{array}
              \right)\,,
\end{equation}
where
\begin{equation}
\sigma_Z = \left(
             \begin{array}{cc}
               1 & 0 \\
               0 & -1 \\
             \end{array}
           \right)\,.
\end{equation}
The symplectic eigenvalues of $\sigma(\omega_{AR}(\eta))$ are
\begin{equation}
\nu_+(\sigma(\omega_{AR}(\eta))) = \left(1-\eta\right)E + \tfrac{1}{2}\,,\qquad \nu_-(\sigma(\omega_{AR}(\eta))) = \tfrac{1}{2}\,,
\end{equation}
hence
\begin{align}
S(ABR)_{\rho_{ABR}(\eta)} &= S(AR)_{\omega_{AR}(\eta)} = g\left(\nu_+(\sigma(\omega_{AR}(\eta)))-\tfrac{1}{2}\right) + g\left(\nu_-(\sigma(\omega_{AR}(\eta)))-\tfrac{1}{2}\right)\nonumber\\
&= g((1-\eta)E)\,,\nonumber\\
S(R)_{\rho_{ABR}(\eta)} &= S(R)_{\omega_{AR}(\eta)} = g(\eta\,E)\,.
\end{align}
Let $\rho_{AR}(\eta)$ and $\rho_{BR}(\eta)$ be the marginals of $\rho_{ABR}(\eta)$ on $AR$ and $BR$, respectively.
They are the quantum Gaussian states with covariance matrices
\begin{align}
\sigma(\rho_{AR}(\eta)) &= \left(
              \begin{array}{cc}
                \left(\kappa\left(E+1\right)-\tfrac{1}{2}\right)I_2 & \sqrt{\kappa\,\eta\,E\left(E+1\right)}\,\sigma_Z \\
                \sqrt{\kappa\,\eta\,E\left(E+1\right)}\,\sigma_Z & \left(\eta\,E + \tfrac{1}{2}\right)I_2 \\
              \end{array}
            \right)\,,\nonumber\\
\sigma(\rho_{BR}(\eta)) &= \left(
              \begin{array}{cc}
                \left(\left(\kappa-1\right)\left(E+\tfrac{1}{2}\right) + \tfrac{\kappa}{2}\right)I_2 & \sqrt{\left(\kappa-1\right)\eta\,E\left(E+1\right)}\,I_2 \\
                \sqrt{\left(\kappa-1\right)\eta\,E\left(E+1\right)}\,I_2 & \left(\eta\,E + \tfrac{1}{2}\right)I_2 \\
              \end{array}
            \right)\,.
\end{align}
Their symplectic eigenvalues are
\begin{align}
\nu_+(\sigma(\rho_{AR}(\eta))) &= \kappa\left(E+1\right) - \eta\,E - \tfrac{1}{2}\,,\nonumber\\
\nu_+(\sigma(\rho_{BR}(\eta))) &= \kappa\left(E+1\right) - \left(1-\eta\right)E - \tfrac{1}{2}\,,\nonumber\\
\nu_-(\sigma(\rho_{AR}(\eta))) &= \nu_-(\sigma(\rho_{BR}(\eta))) = \tfrac{1}{2}\,,
\end{align}
hence
\begin{align}
S(AR)_{\rho_{ABR}(\eta)} &= g\left(\kappa\left(E+1\right) - \eta\,E - 1\right)\,,\nonumber\\
S(BR)_{\rho_{ABR}(\eta)} &= g\left(\kappa\left(E+1\right) - \left(1-\eta\right)E - 1\right)\,.
\end{align}
Therefore, for any $0\le\eta\le1$ we have
\begin{align}\label{eq:Esqg}
E_{\mathrm{sq}}(\rho_{AB}) &\le \tfrac{1}{2}I(A;B|R)_{\rho_{ABR}(\eta)}\nonumber\\
&= \tfrac{1}{2}\left(S(AR)_{\rho_{ABR}(\eta)} + S(BR)_{\rho_{ABR}(\eta)} - S(R)_{\rho_{ABR}(\eta)} - S(ABR)_{\rho_{ABR}(\eta)}\right)\nonumber\\
&= \tfrac{1}{2}\left(\psi_{E,\kappa}(\eta) + \psi_{E,\kappa}(1-\eta)\right)\,,
\end{align}
where
\begin{equation}
\psi_{E,\kappa}(\eta) = g(\kappa\,E+\kappa-\eta\,E-1) - g((1-\eta)E)\,.
\end{equation}
From \autoref{lem:conv} of \autoref{app:B}, $\psi_{E,\kappa}$ is convex, hence from Jensen's inequality the minimum over $0\le\eta\le1$ of the right-hand side of \eqref{eq:Esqg} is achieved by $\eta=\frac{1}{2}$.
Therefore,
\begin{equation}
E_{\mathrm{sq}}(\rho_{AB}) \le \psi_{E,\kappa}\left(\tfrac{1}{2}\right)\,,
\end{equation}
and the claim follows.
We notice that $S(A|R)_{\omega_{AR}(\frac{1}{2})}=0$, which is the same condition that saturates the last inequality of \eqref{eq:ineqI}.

\paragraph{Gap}
Let
\begin{equation}
f(\kappa,E) = g\left(\left(\kappa-\tfrac{1}{2}\right)E+\kappa-1\right)-g\left(\tfrac{E}{2}\right)-\ln\left(2\kappa-1\right)
\end{equation}
be the difference between the upper and lower bound of \eqref{eq:Esqt}.
Since $E\mapsto f(\kappa,E)$ is decreasing,
\begin{equation}
f(\kappa,E) \le f(\kappa,0) = g(\kappa-1)-\ln\left(2\kappa-1\right)\,.
\end{equation}
We have
\begin{equation}
\frac{\partial f}{\partial \kappa}(\kappa,0) = \ln\frac{\kappa}{\kappa-1} - \frac{2}{2\kappa-1} \ge 0\,,
\end{equation}
hence $\kappa\mapsto f(\kappa,0)$ is increasing and
\begin{equation}
f(\kappa,0) \le \lim_{\kappa\to\infty}f(\kappa,0) = \ln\frac{\mathrm{e}}{2}\,.
\end{equation}
\end{proof}
\end{thm}

\begin{figure}
\includegraphics[width=\linewidth]{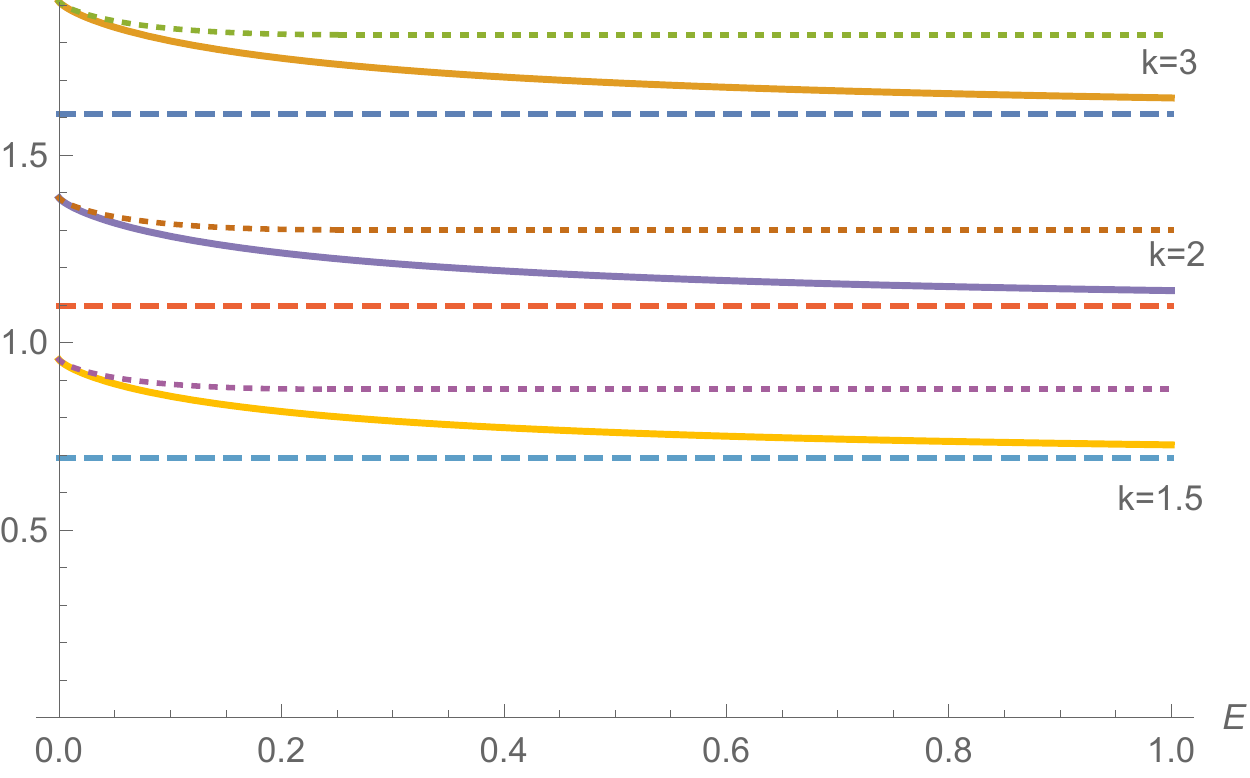}
\caption{Thick and dashed: the upper and lower bounds of \autoref{thm:squashedrho} to the squashed entanglement of the quantum Gaussian state $\rho_{AB}^{\kappa,E}$ of \eqref{eq:defrho} for $\kappa=1.5,\,2,\,3$ and $0\le E\le1$.
The bounds coincide in the limit of infinite average energy, but they are extremely close already from $E\simeq1$.
Dotted: the classical squashed entanglement of $\rho_{AB}^{\kappa,E}$ determined in \autoref{thm:Esqc} for the same values of $\kappa$ and $E$.}
\label{fig:bounds}
\end{figure}

\autoref{fig:bounds} shows the difference between the upper and lower bounds of \eqref{eq:Esqt}, which is extremely small already from $E\simeq1$.

\autoref{lem:equiv} below provides the link between \autoref{thm:squashedrho} and the noiseless quantum Gaussian attenuators and amplifiers.
For any $E\ge0$, let $|\phi_E\rangle_{AB}$ be the two-mode squeezed vacuum state on $AB$ with average energy per mode $E$, which coincides with the quantum Gaussian state $\rho_{AB}^{\kappa',E'}$ of \eqref{eq:defrho} with $\kappa'=E+1$ and $E'=0$.

Let $\gamma_{AB}^{\eta,E}$ be the quantum Gaussian state obtained sending the $B$ system of $|\phi_E\rangle_{AB}$ through a noiseless quantum Gaussian attenuator with attenuation parameter $0\le\eta\le1$, and let $\gamma_{AB}^{\kappa,E}$ be the quantum Gaussian state obtained sending the $A$ system of $|\phi_E\rangle_{AB}$ through a noiseless quantum Gaussian amplifier with amplification parameter $\kappa\ge1$:
\begin{equation}\label{eq:defrho'}
\gamma_{AB}^{\eta,E} = \left(\mathbb{I}_A\otimes\mathcal{E}_\eta\right)(|\phi_E\rangle_{AB}\langle\phi_E|)\,,\qquad \gamma_{AB}^{\kappa,E} = \left(\mathcal{A}_\kappa\otimes\mathbb{I}_B\right)(|\phi_E\rangle_{AB}\langle\phi_E|)\,.
\end{equation}
\begin{cor}\label{lem:equiv}
The squashed entanglement of the quantum Gaussian states $\gamma_{AB}^{\eta,E}$ and $\gamma_{AB}^{\kappa,E}$ defined in \eqref{eq:defrho'} satisfies
\begin{align}\label{eq:Esqch}
\ln\frac{\left(1+\eta\right)E+1}{\left(1-\eta\right)E+1} & \le E_{\mathrm{sq}}\left(\gamma_{AB}^{\eta,E}\right) \le g\left(\tfrac{1+\eta}{2}\,E\right) - g\left(\tfrac{1-\eta}{2}\,E\right)\,,\nonumber\\
\ln\frac{\left(\kappa+1\right)E+\kappa}{\left(\kappa-1\right)E+\kappa} & \le E_{\mathrm{sq}}\left(\gamma_{AB}^{\kappa,E}\right) \le g\left(\frac{\left(\kappa+1\right)E+\kappa-1}{2}\right) - g\left(\tfrac{\kappa-1}{2}\left(E+1\right)\right)\,.
\end{align}
Moreover, the gap between the upper bounds and the respective lower bounds in \eqref{eq:Esqch} is at most $\ln\frac{\mathrm{e}}{2}\simeq0.31$.
We conjecture that the actual value of the squashed entanglement of $\gamma_{AB}^{\eta,E}$ and $\gamma_{AB}^{\kappa,E}$ coincides with the upper bounds in \eqref{eq:Esqch}.
\begin{proof}
We will prove that both the lower and the upper bounds of \eqref{eq:Esqch} follow from \autoref{thm:squashedrho}.
While the lower bounds appear in this paper for the first time, the upper bounds have been proved also in \onlinecite{takeoka2014fundamental,takeoka2014squashed,goodenough2016assessing,davis2018energy}.

Let $\rho_{AB}^{\kappa',E'}$ be as in \eqref{eq:defrho} with $\kappa'\ge1$ and $E'\ge0$.
The covariance matrices of $\rho_{AB}^{\kappa',E'}$, $\gamma_{AB}^{\eta,E}$ and $\gamma_{AB}^{\kappa,E}$ are
\begin{align}\label{eq:sigma}
\sigma\left(\rho_{AB}^{\kappa',E'}\right) &= \left(
                \begin{array}{cc}
                  \left(\kappa'\left(E'+1\right)-\tfrac{1}{2}\right)I_2 & \left(E'+1\right)\sqrt{\kappa'\left(\kappa'-1\right)}\,\sigma_Z \\
                  \left(E'+1\right)\sqrt{\kappa'\left(\kappa'-1\right)}\,\sigma_Z & \left(\left(\kappa'-1\right)\left(E'+1\right)+\tfrac{1}{2}\right)I_2\\
                \end{array}
              \right)\,,\nonumber\\
\sigma\left(\gamma_{AB}^{\eta,E}\right) &= \left(
                 \begin{array}{cc}
                   \left(E+\tfrac{1}{2}\right)I_2 & \sqrt{\eta\,E\left(E+1\right)}\,\sigma_Z \\
                   \sqrt{\eta\,E\left(E+1\right)}\,\sigma_Z & \left(\eta\,E + \tfrac{1}{2}\right)I_2 \\
                 \end{array}
               \right)\,,\nonumber\\
\sigma\left(\gamma_{AB}^{\kappa,E}\right) &= \left(
                         \begin{array}{cc}
                           \left(\kappa\,E+\kappa-\tfrac{1}{2}\right)I_2 & \sqrt{\kappa\,E\left(E+1\right)}\,\sigma_Z \\
                           \sqrt{\kappa\,E\left(E+1\right)}\,\sigma_Z & \left(E+\tfrac{1}{2}\right)I_2 \\
                         \end{array}
                       \right)\,.
\end{align}
For
\begin{equation}
\kappa' = \frac{E+1}{\left(1-\eta\right)E+1}\,,\qquad E' = \left(1-\eta\right)E
\end{equation}
we have $\sigma\left(\rho_{AB}^{\kappa',E'}\right) = \sigma\left(\gamma_{AB}^{\eta,E}\right)$, and therefore $\rho_{AB}^{\kappa',E'}=\gamma_{AB}^{\eta,E}$.
Analogously, for
\begin{equation}
\kappa' = \frac{\kappa\left(E+1\right)}{\left(\kappa-1\right)E+\kappa}\,,\qquad E'=\left(\kappa-1\right)\left(E+1\right)
\end{equation}
we have $\sigma\left(\rho_{AB}^{\kappa',E'}\right) = \sigma\left(\gamma_{AB}^{\kappa,E}\right)$, and therefore $\rho_{AB}^{\kappa',E'}=\gamma_{AB}^{\kappa,E}$.
The claim then follows from \autoref{thm:squashedrho}.
\end{proof}
\begin{rem}
Since the squashed entanglement of any bipartite quantum state is lower than the squashed entanglement of the quantum Gaussian state with the same covariance matrix \cite{wolf2006extremality}, the upper bounds of \eqref{eq:Esqt} and \eqref{eq:Esqch} apply to any bipartite quantum state with covariance matrix as in \eqref{eq:sigma}.
\end{rem}
\end{cor}
We can now prove the main result of the paper.
\begin{thm}\label{thm:main}
The squashed entanglement of the noiseless quantum Gaussian attenuator with attenuation parameter $0\le\eta\le1$ and of the noiseless quantum Gaussian amplifier with amplification parameter $\kappa\ge1$ is
\begin{equation}
E_{\mathrm{sq}}(\mathcal{E}_\eta) = \ln\frac{1+\eta}{1-\eta}\,,\qquad E_{\mathrm{sq}}(\mathcal{A}_\kappa) = \ln\frac{\kappa+1}{\kappa-1}\,,
\end{equation}
and is asymptotically achieved sending half of a two-mode squeezed vacuum state $|\phi_E\rangle_{AB}$ in the limit $E\to\infty$ of infinite average energy or infinite squeezing.
\begin{proof}
The upper bounds
\begin{equation}
E_{\mathrm{sq}}(\mathcal{E}_\eta) \le \ln\frac{1+\eta}{1-\eta}\,,\qquad E_{\mathrm{sq}}(\mathcal{A}_\kappa) \le \ln\frac{\kappa+1}{\kappa-1}
\end{equation}
have been proved in \onlinecite{takeoka2014fundamental,takeoka2014squashed,goodenough2016assessing,davis2018energy}.

Given $E\ge0$, let $\gamma_{AB}^{\eta,E}$ and $\gamma_{AB}^{\kappa,E}$ be as in \eqref{eq:defrho'}.
We have from \autoref{lem:equiv}
\begin{align}
E_{\mathrm{sq}}(\mathcal{E}_\eta) &\ge E_{\mathrm{sq}}\left(\gamma_{AB}^{\eta,E}\right) \ge \ln\frac{\left(1+\eta\right)E+1}{\left(1-\eta\right)E+1}\,,\nonumber\\
E_{\mathrm{sq}}(\mathcal{A}_\kappa) &\ge E_{\mathrm{sq}}\left(\gamma_{AB}^{\kappa,E}\right) \ge \ln\frac{\left(\kappa+1\right)E+\kappa}{\left(\kappa-1\right)E+\kappa}\,.
\end{align}
Taking the limit $E\to\infty$ we get
\begin{equation}
E_{\mathrm{sq}}(\mathcal{E}_\eta) \ge \ln\frac{1+\eta}{1-\eta}\,,\qquad E_{\mathrm{sq}}(\mathcal{A}_\kappa) \ge \ln\frac{\kappa+1}{\kappa-1}\,,
\end{equation}
and the claim follows.
\end{proof}
\end{thm}

\section{Classical squashed entanglement}\label{sec:Esqc}
The classical squashed entanglement \cite{yang2007conditional,yang2009squashed,song2009lower,li2014relative,huang2014computing} has the same definition as the squashed entanglement with the minimization in \eqref{eq:defEsq} restricted to the classical extensions of the quantum state.
A classical extension $\rho_{ABR}$ of the quantum state $\rho_{AB}$ is given by a probability measure $\rho_R$ on a measure space $\mathcal{R}$ and a set $\{\rho_{AB|R=r}\}_{r\in\mathcal{R}}$ of the states of the quantum system $AB$ conditioned on $R=r$, such that the function $r\mapsto\rho_{AB|R=r}$ is measurable and
\begin{equation}
\int_{\mathcal{R}}\rho_{AB|R=r}\,\mathrm{d}\rho_R(r)=\rho_{AB}\,.
\end{equation}
The classical squashed entanglement of the bipartite quantum state $\rho_{AB}$ is half of the infimum over all the classical extension $\rho_{ABR}$ of the mutual information between the quantum systems $A$ and $B$ conditioned on the classical system $R$:
\begin{equation}\label{eq:defEsqc}
E_{\mathrm{sq,c}}(\rho_{AB}) = \tfrac{1}{2}\inf\left\{I(A;B|R)_{\rho_{ABR}}:\int_{\mathcal{R}}\rho_{AB|R=r}\,\mathrm{d}\rho_R(r)=\rho_{AB}\right\}\,,
\end{equation}
where the conditional mutual information is defined as
\begin{equation}
I(A;B|R)_{\rho_{ABR}} = \int_{\mathcal{R}}I(A;B)_{\rho_{AB|R=r}}\,\mathrm{d}\rho_R(r)\,.
\end{equation}
The classical squashed entanglement is always not lower than the squashed entanglement, and can be strictly larger \cite{brandao2008entanglement}.
A fundamental property of the squashed entanglement is its additivity with respect to the tensor product \cite{christandl2004squashed}: for any two bipartite quantum states $\rho_{A_1B_1}$ and $\rho_{A_2B_2}$,
\begin{equation}
E_{\mathrm{sq}}(\rho_{A_1B_1}\otimes\rho_{A_2B_2}) = E_{\mathrm{sq}}(\rho_{A_1B_1}) + E_{\mathrm{sq}}(\rho_{A_2B_2})\,.
\end{equation}
The classical squashed entanglement is subadditive with respect to the tensor product:
\begin{equation}
E_{\mathrm{sq,c}}(\rho_{A_1B_1}\otimes\rho_{A_2B_2}) \le E_{\mathrm{sq,c}}(\rho_{A_1B_1}) + E_{\mathrm{sq,c}}(\rho_{A_2B_2})\,,
\end{equation}
but it is not known whether it is additive.
This has led to the definition of the asymptotic classical squashed entanglement as the regularization of the classical squashed entanglement over many copies of the quantum state \cite{yang2007conditional,yang2009squashed,li2014relative}:
\begin{equation}
E_{\mathrm{sq,c}}^\infty(\rho_{AB}) = \lim_{n\to\infty}\frac{E_{\mathrm{sq,c}}\left(\rho_{AB}^{\otimes n}\right)}{n}\,.
\end{equation}
Thanks to the additivity of the squashed entanglement, the asymptotic classical squashed entanglement is still an upper bound to the squashed entanglement.

Here we determine the classical squashed entanglement of the Gaussian quantum states defined in \eqref{eq:defrho} and show that it is achieved by a Gaussian extension and strictly larger than their squashed entanglement (see \autoref{fig:bounds} for the comparison).
The proof exploits the constrained minimum output entropy conjecture for the one-mode noiseless quantum Gaussian amplifier and its complementary channel.
We also show that the multi-mode generalization of the conjecture determines the asymptotic classical squashed entanglement of the Gaussian quantum states \eqref{eq:defrho} and implies that it is equal to the classical squashed entanglement of one copy of the state, and therefore still strictly larger than the squashed entanglement.

\begin{thm}\label{thm:Esqc}
The classical squashed entanglement of the Gaussian quantum state \eqref{eq:defrho} is achieved by a Gaussian extension and is equal to
\begin{equation}\label{eq:Esq}
E_{\mathrm{sq,c}}\left(\rho_{AB}^{\kappa,E}\right) = \frac{1}{2}\min_{x\in[0,E]}h_\kappa(x)\,,
\end{equation}
where for any $x\ge0$,
\begin{equation}
h_\kappa(x) = g(\kappa\,x + \kappa - 1) + g((\kappa-1)(x+1)) - g(x)\,.
\end{equation}
For any $E>0$ and any $\kappa>1$, the classical squashed entanglement of $\rho_{AB}^{\kappa,E}$ is strictly larger than its squashed entanglement:
\begin{equation}
E_{\mathrm{sq,c}}\left(\rho_{AB}^{\kappa,E}\right) > E_{\mathrm{sq}}\left(\rho_{AB}^{\kappa,E}\right)\,.
\end{equation}
Moreover, assuming \autoref{conj:MOE}, the asymptotic classical squashed entanglement of $\rho_{AB}^{\kappa,E}$ coincides with its classical squashed entanglement:
\begin{equation}
E_{\mathrm{sq,c}}^\infty\left(\rho_{AB}^{\kappa,E}\right) = E_{\mathrm{sq,c}}\left(\rho_{AB}^{\kappa,E}\right)\,.
\end{equation}

\begin{rem}\label{rem}
Since the function $s\mapsto h_\kappa(g^{-1}(s))$ is strictly convex [\onlinecite[Lemma 15]{de2017multimode}] and
\begin{equation}
\left.\frac{\mathrm{d}}{\mathrm{d}s}h_\kappa(g^{-1}(s))\right|_{s=0} = -\frac{1}{2}\,,\qquad \lim_{s\to\infty}\frac{\mathrm{d}}{\mathrm{d}s}h_\kappa(g^{-1}(s)) = \frac{1}{2}\,,
\end{equation}
$s\mapsto h_\kappa(g^{-1}(s))$ has a unique local minimum, which is also the global minimum.
If $s_\kappa$ is the minimizer, $h_\kappa(g^{-1}(s))$ is strictly decreasing for $0\le s\le s_\kappa$ and strictly increasing for $s\ge s_\kappa$.
Since $g$ is strictly increasing, $h_{\kappa}(x)$ attains its global minimum in $x = E_\kappa = g^{-1}(s_\kappa)$, is strictly decreasing for $0\le x\le E_\kappa$ and strictly increasing for $x\ge E_\kappa$.
Therefore,
\begin{equation}
\min_{x\in[0,E]}h_\kappa(x) = \left\{
                                  \begin{array}{ll}
                                    h_\kappa(E)\,, & 0\le E\le E_\kappa \\
                                    h_\kappa(E_\kappa)\,, & E\ge E_\kappa \\
                                  \end{array}
                                \right.\,.
\end{equation}
\end{rem}

\begin{proof}
Let $\rho_{A_1^nB_1^nR}$ be a classical extension of $\left(\rho_{AB}^{\kappa,E}\right)^{\otimes n}$, where $A_1^n=A_1\ldots A_n$ and analogously for $B_1^n$.
The quantum state $U_{\kappa}^{\dag\otimes n}\,\rho_{A_1^nB_1^nR}\,U_{\kappa}^{\otimes n}$ is a classical extension of ${\omega_A(E)}^{\otimes n}\otimes{|0\rangle_B\langle0|}^{\otimes n}$, therefore it has the form $\omega_{A_1^nR}\otimes{|0\rangle_B\langle0|}^{\otimes n}$ for some quantum-classical state $\omega_{A_1^nR}$ that is a classical extension of ${\omega_A(E)}^{\otimes n}$.
The quantum-classical state $\rho_{A_1^nB_1^nR}$ has then the form
\begin{equation}\label{eq:rhoABRc}
\rho_{A_1^nB_1^nR} = U_{\kappa}^{\otimes n}\left(\omega_{A_1^nR}\otimes{|0\rangle_B\langle0|}^{\otimes n}\right)U_{\kappa}^{\dag\otimes n}\,,\quad\int_{\mathcal{R}}\omega_{A_1^n|R=r}\,\mathrm{d}\rho_R(r)={\omega_A(E)}^{\otimes n}\,.
\end{equation}
Conversely, any state of the form \eqref{eq:rhoABRc} is a classical extension of $\rho_{AB}^{\kappa,E}$.
We have for any $r\in\mathcal{R}$
\begin{align}
\rho_{A_1^nB_1^n|R=r} &= U_{\kappa}^{\otimes n}\left(\omega_{A_1^n|R=r}\otimes{|0\rangle_B\langle0|}^{\otimes n}\right)U_{\kappa}^{\dag\otimes n}\,,\nonumber\\
\mathrm{Tr}_{B_1^n}\,\rho_{A_1^nB_1^n|R=r} &= \mathcal{A}_{\kappa}^{\otimes n}(\omega_{A_1^n|R=r})\,,\nonumber\\
\mathrm{Tr}_{A_1^n}\,\rho_{A_1^nB_1^n|R=r} &= \tilde{\mathcal{A}}_{\kappa}^{\otimes n}(\omega_{A_1^n|R=r})\,,
\end{align}
where $\tilde{\mathcal{A}}_{\kappa}$ is the complementary channel of $\mathcal{A}_\kappa$ \cite{holevo2013quantum}.
Then,
\begin{align}\label{eq:Esqc}
&I(A_1^n;B_1^n|R)_{\rho_{A_1^nB_1^nR}} = \int_{\mathcal{R}}I(A_1^n;B_1^n)_{\rho_{A_1^nB_1^n|R=r}}\,\mathrm{d}\rho_R(r)\nonumber\\
&= \int_{\mathcal{R}}\left(S\left(\mathcal{A}_{\kappa}^{\otimes n}(\omega_{A_1^n|R=r})\right) + S\left(\tilde{\mathcal{A}}_{\kappa}^{\otimes n}(\omega_{A_1^n|R=r})\right) - S(\omega_{A_1^n|R=r})\right)\mathrm{d}\rho_R(r)\,.
\end{align}

\paragraph{Upper bound}
Let $n=1$ and let $E'\in[0,E]$.
Choosing $\mathcal{R}=\mathbb{C}$, we have
\begin{equation}
\omega(E) = \int_{\mathbb{C}}D_r\,\omega(E')\,D_r^\dag\,\mathrm{e}^{-\frac{|r|^2}{E-E'}}\,\frac{\mathrm{d}r}{\pi}\,,
\end{equation}
where for any $r\in\mathbb{C}$, $D_r$ is the unitary operator that displaces by $r$ the ladder operator \cite{holevo2013quantum}:
\begin{equation}
D_r^\dag\,a\,D_r = a + r\,\mathbb{I}\,.
\end{equation}
We can then choose $\omega_{A|R=r} = \omega(E')$ for any $r\in\mathbb{C}$.
Since
\begin{equation}
\mathcal{A}_\kappa(\omega(E')) = \omega(\kappa\,E' + \kappa - 1)\,,\qquad \tilde{\mathcal{A}}_\kappa(\omega(E')) = \omega((\kappa-1)(E'+1))\,,
\end{equation}
we have from \eqref{eq:Esqc}
\begin{equation}\label{eq:Esqci}
E_{\mathrm{sq,c}}^\infty\left(\rho_{AB}^{\kappa,E}\right) \le E_{\mathrm{sq,c}}\left(\rho_{AB}^{\kappa,E}\right) \le \tfrac{1}{2}I(A;B|R)_{\rho_{ABR}} = \tfrac{1}{2}h_\kappa(E')\,,
\end{equation}
and the claim follows taking the minimum of the right-hand side of \eqref{eq:Esqci} over $E'\in[0,E]$.

\paragraph{Lower bound}
$\mathbf{E_{\mathrm{sq,c}}}:$
Let $n=1$.
\autoref{thm:1MOE} and \autoref{thm:MOE} imply for any $r\in\mathcal{R}$
\begin{equation}\label{eq:MOE1}
S\left(\mathcal{A}_{\kappa}(\omega_{A|R=r})\right) + S\left(\tilde{\mathcal{A}}_{\kappa}(\omega_{A|R=r})\right) - S(\omega_{A|R=r}) \ge h_\kappa(g^{-1}(S(\omega_{A|R=r})))\,.
\end{equation}
From [\onlinecite[Lemma 15]{de2017multimode}], the function $s\mapsto h_\kappa(g^{-1}(s))$ is convex.
We then have from \eqref{eq:Esqc}, \eqref{eq:MOE1} and Jensen's inequality
\begin{align}\label{eq:Esqcf1}
I(A;B|R)_{\rho_{ABR}} &\ge \int_{\mathcal{R}}h_\kappa(g^{-1}(S(\omega_{A|R=r})))\,\mathrm{d}\rho_R(r) \ge h_\kappa(g^{-1}(S(A|R)_{\omega_{AR}}))\nonumber\\
&\ge \inf_{E'\in[0,E]}h_\kappa(E')\,,
\end{align}
where we have used that $0\le g^{-1}(S(A|R)_{\omega_{AR}})\le E$ since $g^{-1}$ is increasing and
\begin{equation}
0\le \int_{\mathcal{R}}S(\omega_{A|R=r})\,\mathrm{d}\rho_R(r) = S(A|R)_{\omega_{AR}} \le S(A)_{\omega_{AR}} = g(E)\,.
\end{equation}
Finally, taking the infimum of the left-hand side of \eqref{eq:Esqcf1} over all the classical extensions $\rho_{ABR}$ of $\rho_{AB}^{\kappa,E}$ we get
\begin{equation}
E_{\mathrm{sq,c}}\left(\rho_{AB}^{\kappa,E}\right) \ge \tfrac{1}{2}\inf_{E'\in[0,E]}h_\kappa(E')\,.
\end{equation}

$\mathbf{E_{\mathrm{sq,c}}^\infty}:$
The proof with generic $n$ is analogous to the proof for $n=1$: \autoref{conj:MOE} and \autoref{thm:MOE} imply for any $r\in\mathcal{R}$
\begin{align}\label{eq:MOE}
&S\left(\mathcal{A}_{\kappa}^{\otimes n}(\omega_{A_1^n|R=r})\right) + S\left(\tilde{\mathcal{A}}_{\kappa}^{\otimes n}(\omega_{A_1^n|R=r})\right) - S(\omega_{A_1^n|R=r})\nonumber\\
&\ge n\,h_\kappa\left(g^{-1}\left(\tfrac{1}{n}S(\omega_{A_1^n|R=r})\right)\right)\,.
\end{align}
We then have from \eqref{eq:Esqc}, \eqref{eq:MOE} and Jensen's inequality
\begin{align}\label{eq:Esqcf}
I(A_1^n;B_1^n|R)_{\rho_{A_1^nB_1^nR}} &\ge n\int_{\mathcal{R}}h_\kappa\left(g^{-1}\left(\tfrac{1}{n}S(\omega_{A_1^n|R=r})\right)\right)\,\mathrm{d}\rho_R(r)\nonumber\\
&\ge n\,h_\kappa\left(g^{-1}\left(\tfrac{1}{n}S(A_1^n|R)_{\omega_{A_1^nR}}\right)\right) \ge n\inf_{E'\in[0,E]}h_\kappa(E')\,,
\end{align}
where we have used that $0\le g^{-1}\left(\tfrac{1}{n}S(A_1^n|R)_{\omega_{A_1^nR}}\right)\le E$ since $g^{-1}$ is increasing and
\begin{equation}
0\le \int_{\mathcal{R}}S(\omega_{A_1^n|R=r})\,\mathrm{d}\rho_R(r) = S(A_1^n|R)_{\omega_{A_1^nR}} \le S(A_1^n)_{\omega_{A_1^nR}} = n\,g(E)\,.
\end{equation}
Taking the infimum of the left-hand side of \eqref{eq:Esqcf} over all the classical extensions $\rho_{A_1^nB_1^nR}$ of $\rho_{A_1^nB_1^n}^{\kappa,E}$ we get
\begin{equation}
E_{\mathrm{sq,c}}\left({\rho_{AB}^{\kappa,E}}^{\otimes n}\right) \ge \frac{n}{2}\inf_{E'\in[0,E]}h_\kappa(E')\,,
\end{equation}
and finally
\begin{equation}
E_{\mathrm{sq,c}}^\infty\left(\rho_{AB}^{\kappa,E}\right) = \lim_{n\to\infty}\frac{1}{n}E_{\mathrm{sq,c}}\left({\rho_{AB}^{\kappa,E}}^{\otimes n}\right) \ge \tfrac{1}{2}\inf_{E'\in[0,E]}h_\kappa(E')\,.
\end{equation}

\paragraph{Separation between squashed entanglement and classical squashed entanglement}
From the upper bound of \eqref{eq:Esqt} and \autoref{rem}, it is sufficient to prove that for any $\kappa>1$,
\begin{equation}
g\left(\left(\kappa-\tfrac{1}{2}\right)E+\kappa-1\right)-g\left(\tfrac{E}{2}\right) < \tfrac{1}{2}\left\{
                                                                                                      \begin{array}{cc}
                                                                                                        h_\kappa(E)\,, & 0<E\le E_\kappa \\
                                                                                                        h_\kappa(E_\kappa)\,, & E\ge E_\kappa \\
                                                                                                      \end{array}
                                                                                                    \right.\,.
\end{equation}
Let us consider the case $0<E\le E_\kappa$.
The claim is equivalent to
\begin{equation}\label{eq:kappa}
2g((\kappa-\tfrac{1}{2})E+\kappa-1) - g(\kappa\,E+\kappa-1) - g((\kappa-1)(E+1)) < 2g(\tfrac{E}{2}) - g(E)\,.
\end{equation}
For $\kappa=1$, equality holds in \eqref{eq:kappa}.
The derivative with respect to $\kappa$ of the left-hand side of \eqref{eq:kappa} is
\begin{equation*}
2(E+1)\left(g'((\kappa-\tfrac{1}{2})E+\kappa-1) - \frac{g'(\kappa\,E+\kappa-1) + g'((\kappa-1)(E+1))}{2}\right),
\end{equation*}
and is strictly negative since $g'$ is strictly convex, hence the claim follows.

For $E\ge E_\kappa$, the claim follows since
\begin{align}
g\left(\left(\kappa-\tfrac{1}{2}\right)E+\kappa-1\right)-g\left(\tfrac{E}{2}\right) &\le g\left(\left(\kappa-\tfrac{1}{2}\right)E_\kappa+\kappa-1\right)-g\left(\tfrac{E_\kappa}{2}\right)\nonumber\\
&< \tfrac{1}{2}h_\kappa(E_\kappa)\,.
\end{align}
\end{proof}
\end{thm}

\section{Conclusions and open problems}\label{sec:concl}
We have determined the maximum squashed entanglement achievable between sender and receiver of the noiseless quantum Gaussian attenuator and amplifier (\autoref{thm:main}), and proved that it is strictly larger than the corresponding secret key capacity.
This result opens the way to determine the squashed entanglement of the noisy quantum Gaussian attenuators and amplifiers, for which only upper bounds are known \cite{takeoka2014fundamental,takeoka2014squashed,goodenough2016assessing,davis2018energy}.

Our proof is based on a new lower bound to the squashed entanglement of the quantum Gaussian state \eqref{eq:defrho} obtained applying a two-mode squeezing operation to a quantum thermal Gaussian state tensored with the vacuum state (\autoref{thm:squashedrho}).
Despite being extremely tight, the lower bound is optimal only in the limit of infinite average energy.
Therefore, determining the exact value of the squashed entanglement of the quantum Gaussian state \eqref{eq:defrho} for finite average energy is still an open problem.
We conjecture that this squashed entanglement coincides with the upper bound of \autoref{thm:squashedrho}, which is achieved by a Gaussian extension of the state.
Furthermore, we conjecture that the squashed entanglement of any quantum Gaussian state is achieved by a Gaussian extension.

We have also determined the classical squashed entanglement of the quantum Gaussian state \eqref{eq:defrho}, and proved that it is achieved by a classical Gaussian extension of the state.
This is the first time that the classical squashed entanglement of a quantum Gaussian state is determined.
Therefore, our result opens the way to determine the classical squashed entanglement of all quantum Gaussian states, which we conjecture to be always achieved by a classical Gaussian extension.

\begin{acknowledgments}
We thank Matthias Christandl for fruitful discussions and Stefano Pirandola and Mark M. Wilde for useful comments.

\includegraphics[width=0.05\textwidth]{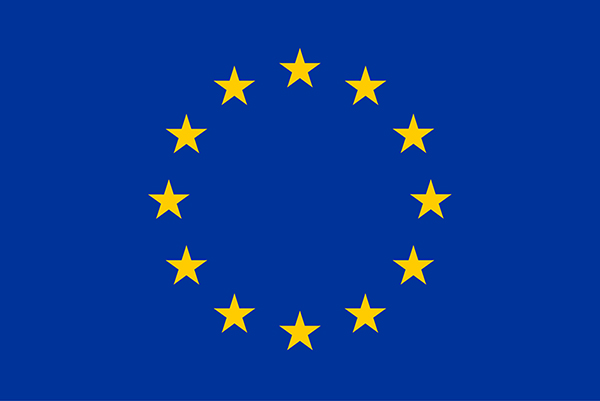}
This project has received funding from the European Union's Horizon 2020 research and innovation programme under the Marie Sk\l odowska-Curie grant agreement No. 792557.

We acknowledge support from the QuantERA ERA-NET Cofund in Quantum Technologies implemented within the European Union's Horizon 2020 Programme (QuantAlgo project) via the Innovation Fund Denmark.

\end{acknowledgments}

\appendix

\section{Entropic inequalities}\label{app}
A longstanding conjecture in quantum communication theory states that pairs of $n$-mode thermal quantum Gaussian states minimize the output entropy of the beam-splitter and the squeezing operation among all the pairs of $n$-mode input states with assigned entropies \cite{de2018gaussian}.
This conjecture was first formulated in 2007 to determine the maximum rates for classical communication to two receivers with the noiseless quantum Gaussian broadcast channel \cite{guha2007classical,guha2007classicalproc,guha2008entropy,guha2008capacity}.
A fundamental step towards the proof of the conjecture has been the proof of the quantum Entropy Power Inequality \cite{konig2014entropy,konig2016corrections,de2014generalization,de2015multimode,de2017gaussian}, which provides an almost optimal lower bound to the output entropy of the beam-splitter and the squeezing.
The constrained minimum output entropy conjecture has then been proved for $n=1$ in the special case when one of the two input states of the beam-splitter or of the squeezing operation is Gaussian, and therefore the quantum channel that maps the input to the output is a quantum Gaussian attenuator, amplifier or phase-contravariant channel \cite{de2015passive,de2016passive,de2016gaussian,qi2017minimum,de2018pq,de2016gaussiannew}.
In this paper we exploit the constrained minimum output entropy conjecture for the noiseless quantum Gaussian amplifier:
\begin{thm}[one-mode constrained minimum output entropy conjecture {\cite{de2016gaussiannew}}]\label{thm:1MOE}
Quantum thermal Gaussian input states minimize the output entropy of the one-mode noiseless quantum Gaussian amplifier among all the input states with a given entropy.
In other words, for any one-mode quantum state $\rho$, let
\begin{equation}
\omega_\rho = \omega\left(g^{-1}(S(\rho))\right)
\end{equation}
be the one-mode quantum thermal Gaussian state as in \eqref{eq:omega} with the same entropy as $\rho$.
Then, for any $\kappa\ge1$
\begin{equation}
S\left(\mathcal{A}_{\kappa}(\rho)\right) \ge S\left(\mathcal{A}_{\kappa}(\omega_\rho)\right) = g\left(\kappa\,g^{-1}(S(\rho)) + \kappa-1\right)\,.
\end{equation}
\end{thm}
For $n\ge2$, the constrained minimum output entropy conjecture is still open in general, and has been proved only for the quantum Gaussian channels that are entanglement breaking \cite{de2017multimode,de2019new}, which include the complementary channel of the noiseless quantum Gaussian amplifier:
\begin{conj}[minimum output entropy conjecture {[\onlinecite[Conjecture V.1]{de2018gaussian}]}]\label{conj:MOE}
For any $n\in\mathbb{N}$, quantum thermal Gaussian input states minimize the output entropy of the $n$-mode Gaussian quantum-limited amplifier among all the input states with a given entropy.
In other words, for any $n$-mode quantum state $\rho$, let
\begin{equation}
\omega_\rho = {\omega\left(g^{-1}\left(\tfrac{1}{n}S(\rho)\right)\right)}^{\otimes n}
\end{equation}
be the $n$-mode quantum thermal Gaussian state with the same entropy as $\rho$.
Then, for any $\kappa\ge1$
\begin{equation}
S\left(\mathcal{A}^{\otimes n}_{\kappa}(\rho)\right) \ge S\left(\mathcal{A}^{\otimes n}_{\kappa}(\omega_\rho)\right) = n\,g\left(\kappa\,g^{-1}\left(\tfrac{1}{n}S(\rho)\right) + \kappa-1\right)\,.
\end{equation}
\end{conj}

\begin{rem}
\autoref{conj:MOE} has been proved in the particular case when the input state $\rho$ is pure \cite{giovannetti2015solution,holevo2015gaussian,giovannetti2015majorization,giovannetti2014ultimate}.
\end{rem}

\begin{thm}[{[\onlinecite[Corollary 5]{de2019new}]}]\label{thm:MOE}
For any $n\in\mathbb{N}$, quantum thermal Gaussian input states minimize the output entropy of the complementary channel of the $n$-mode noiseless quantum Gaussian amplifier among all the input states with a given entropy.
In other words, for any $n$-mode quantum state $\rho$, let
\begin{equation}
\omega_\rho = {\omega\left(g^{-1}\left(\tfrac{1}{n}S(\rho)\right)\right)}^{\otimes n}
\end{equation}
be the $n$-mode quantum thermal Gaussian state with the same entropy as $\rho$.
Then, for any $\kappa\ge1$
\begin{equation}
S\left(\tilde{\mathcal{A}}^{\otimes n}_{\kappa}(\rho)\right) \ge S\left(\tilde{\mathcal{A}}^{\otimes n}_{\kappa}(\omega_\rho)\right) = n\,g\left(\left(\kappa-1\right)\left(g^{-1}\left(\tfrac{1}{n}S(\rho)\right) + 1\right)\right)\,.
\end{equation}
\end{thm}

A conditional version of the quantum Entropy Power Inequality has been proved, where all the entropies are conditioned on an external quantum system \cite{koenig2015conditional,de2018conditional,huber2018conditional}.
In this paper, we exploit its version for the two-mode squeezing operation:
\begin{thm}[quantum conditional Entropy Power Inequality \cite{de2018conditional}]\label{thm:cEPI}
Let $A$ be a one-mode Gaussian quantum system and let $R$ be a generic quantum system.
Let $\gamma_{AR}$ be a joint quantum state on $AR$ such that its marginal $\gamma_A$ on $A$ has finite average energy and its marginal $\gamma_R$ on $R$ has finite entropy.
Let
\begin{equation}
\rho_{ABR} = U_{\kappa}\left(\gamma_{AR}\otimes|0\rangle_B\langle0|\right)U_{\kappa}^\dag\,,
\end{equation}
where $|0\rangle_B$ is the vacuum state of $B$ and $U_{\kappa}$ is the two-mode squeezing operator defined in \eqref{eq:defUk}.
Then,
\begin{align}
S(A|R)_{\rho_{ABR}} &\ge \ln\left(\kappa\exp S(A|R)_{\gamma_{AR}} + \kappa-1\right)\,,\nonumber\\
S(B|R)_{\rho_{ABR}} &\ge \ln\left(\left(\kappa-1\right)\exp S(A|R)_{\gamma_{AR}}+\kappa\right)\,.
\end{align}
\end{thm}

\section{}\label{app:B}
\begin{lem}\label{lem:conv}
For any $E\ge0$ and $\kappa\ge1$, the function $\psi_{E,\kappa}$ is convex.
\begin{proof}
We have
\begin{equation}
\psi_{E,\kappa}''(\eta) = \frac{E\left(E+1\right)\left(\kappa-1\right)\left(\left(\kappa+1-2\eta\right)E+\kappa\right)} {\left(1-\eta\right)\left(\left(\kappa-\eta\right)E+\kappa-1\right)\left(\left(\kappa-\eta\right)E+\kappa\right)\left(\left(1-\eta\right)E+1\right)}\ge0\,.
\end{equation}
\end{proof}
\end{lem}

\bibliographystyle{apsrev4-1}
\bibliography{biblio}

\begin{thebibliography}{91}%
\makeatletter
\providecommand \@ifxundefined [1]{%
 \@ifx{#1\undefined}
}%
\providecommand \@ifnum [1]{%
 \ifnum #1\expandafter \@firstoftwo
 \else \expandafter \@secondoftwo
 \fi
}%
\providecommand \@ifx [1]{%
 \ifx #1\expandafter \@firstoftwo
 \else \expandafter \@secondoftwo
 \fi
}%
\providecommand \natexlab [1]{#1}%
\providecommand \enquote  [1]{``#1''}%
\providecommand \bibnamefont  [1]{#1}%
\providecommand \bibfnamefont [1]{#1}%
\providecommand \citenamefont [1]{#1}%
\providecommand \href@noop [0]{\@secondoftwo}%
\providecommand \href [0]{\begingroup \@sanitize@url \@href}%
\providecommand \@href[1]{\@@startlink{#1}\@@href}%
\providecommand \@@href[1]{\endgroup#1\@@endlink}%
\providecommand \@sanitize@url [0]{\catcode `\\12\catcode `\$12\catcode
  `\&12\catcode `\#12\catcode `\^12\catcode `\_12\catcode `\%12\relax}%
\providecommand \@@startlink[1]{}%
\providecommand \@@endlink[0]{}%
\providecommand \url  [0]{\begingroup\@sanitize@url \@url }%
\providecommand \@url [1]{\endgroup\@href {#1}{\urlprefix }}%
\providecommand \urlprefix  [0]{URL }%
\providecommand \Eprint [0]{\href }%
\providecommand \doibase [0]{http://dx.doi.org/}%
\providecommand \selectlanguage [0]{\@gobble}%
\providecommand \bibinfo  [0]{\@secondoftwo}%
\providecommand \bibfield  [0]{\@secondoftwo}%
\providecommand \translation [1]{[#1]}%
\providecommand \BibitemOpen [0]{}%
\providecommand \bibitemStop [0]{}%
\providecommand \bibitemNoStop [0]{.\EOS\space}%
\providecommand \EOS [0]{\spacefactor3000\relax}%
\providecommand \BibitemShut  [1]{\csname bibitem#1\endcsname}%
\let\auto@bib@innerbib\@empty
\bibitem [{\citenamefont {Tucci}(1999)}]{tucci1999quantum}%
  \BibitemOpen
  \bibfield  {author} {\bibinfo {author} {\bibfnamefont {R.~R.}\ \bibnamefont
  {Tucci}},\ }\href@noop {} {\bibfield  {journal} {\bibinfo  {journal} {arXiv
  preprint quant-ph/9909041}\ } (\bibinfo {year} {1999})}\BibitemShut {NoStop}%
\bibitem [{\citenamefont {Tucci}(2000{\natexlab{a}})}]{tucci2000separability}%
  \BibitemOpen
  \bibfield  {author} {\bibinfo {author} {\bibfnamefont {R.~R.}\ \bibnamefont
  {Tucci}},\ }\href@noop {} {\bibfield  {journal} {\bibinfo  {journal} {arXiv
  preprint quant-ph/0005119}\ } (\bibinfo {year}
  {2000}{\natexlab{a}})}\BibitemShut {NoStop}%
\bibitem [{\citenamefont {Tucci}(2000{\natexlab{b}})}]{tucci2000entanglement}%
  \BibitemOpen
  \bibfield  {author} {\bibinfo {author} {\bibfnamefont {R.~R.}\ \bibnamefont
  {Tucci}},\ }\href@noop {} {\bibfield  {journal} {\bibinfo  {journal} {arXiv
  preprint quant-ph/0010041}\ } (\bibinfo {year}
  {2000}{\natexlab{b}})}\BibitemShut {NoStop}%
\bibitem [{\citenamefont {Tucci}(2001{\natexlab{a}})}]{tucci2001relaxation}%
  \BibitemOpen
  \bibfield  {author} {\bibinfo {author} {\bibfnamefont {R.~R.}\ \bibnamefont
  {Tucci}},\ }\href@noop {} {\bibfield  {journal} {\bibinfo  {journal} {arXiv
  preprint quant-ph/0101123}\ } (\bibinfo {year}
  {2001}{\natexlab{a}})}\BibitemShut {NoStop}%
\bibitem [{\citenamefont {Tucci}(2001{\natexlab{b}})}]{tucci2001entanglement}%
  \BibitemOpen
  \bibfield  {author} {\bibinfo {author} {\bibfnamefont {R.~R.}\ \bibnamefont
  {Tucci}},\ }\href@noop {} {\bibfield  {journal} {\bibinfo  {journal} {arXiv
  preprint quant-ph/0103040}\ } (\bibinfo {year}
  {2001}{\natexlab{b}})}\BibitemShut {NoStop}%
\bibitem [{\citenamefont {Tucci}(2002)}]{tucci2002entanglement}%
  \BibitemOpen
  \bibfield  {author} {\bibinfo {author} {\bibfnamefont {R.~R.}\ \bibnamefont
  {Tucci}},\ }\href@noop {} {\bibfield  {journal} {\bibinfo  {journal} {arXiv
  preprint quant-ph/0202144}\ } (\bibinfo {year} {2002})}\BibitemShut {NoStop}%
\bibitem [{\citenamefont {Christandl}\ and\ \citenamefont
  {Winter}(2004)}]{christandl2004squashed}%
  \BibitemOpen
  \bibfield  {author} {\bibinfo {author} {\bibfnamefont {M.}~\bibnamefont
  {Christandl}}\ and\ \bibinfo {author} {\bibfnamefont {A.}~\bibnamefont
  {Winter}},\ }\href@noop {} {\bibfield  {journal} {\bibinfo  {journal}
  {Journal of mathematical physics}\ }\textbf {\bibinfo {volume} {45}},\
  \bibinfo {pages} {829} (\bibinfo {year} {2004})}\BibitemShut {NoStop}%
\bibitem [{\citenamefont {Brandao}\ \emph {et~al.}(2011)\citenamefont
  {Brandao}, \citenamefont {Christandl},\ and\ \citenamefont
  {Yard}}]{brandao2011faithful}%
  \BibitemOpen
  \bibfield  {author} {\bibinfo {author} {\bibfnamefont {F.~G.}\ \bibnamefont
  {Brandao}}, \bibinfo {author} {\bibfnamefont {M.}~\bibnamefont {Christandl}},
  \ and\ \bibinfo {author} {\bibfnamefont {J.}~\bibnamefont {Yard}},\
  }\href@noop {} {\bibfield  {journal} {\bibinfo  {journal} {Communications in
  Mathematical Physics}\ }\textbf {\bibinfo {volume} {306}},\ \bibinfo {pages}
  {805} (\bibinfo {year} {2011})}\BibitemShut {NoStop}%
\bibitem [{\citenamefont {Seshadreesan}\ \emph {et~al.}(2015)\citenamefont
  {Seshadreesan}, \citenamefont {Berta},\ and\ \citenamefont
  {Wilde}}]{seshadreesan2015renyi}%
  \BibitemOpen
  \bibfield  {author} {\bibinfo {author} {\bibfnamefont {K.~P.}\ \bibnamefont
  {Seshadreesan}}, \bibinfo {author} {\bibfnamefont {M.}~\bibnamefont {Berta}},
  \ and\ \bibinfo {author} {\bibfnamefont {M.~M.}\ \bibnamefont {Wilde}},\
  }\href@noop {} {\bibfield  {journal} {\bibinfo  {journal} {Journal of Physics
  A: Mathematical and Theoretical}\ }\textbf {\bibinfo {volume} {48}},\
  \bibinfo {pages} {395303} (\bibinfo {year} {2015})}\BibitemShut {NoStop}%
\bibitem [{\citenamefont {Wilde}(2017)}]{wilde2017quantum}%
  \BibitemOpen
  \bibfield  {author} {\bibinfo {author} {\bibfnamefont {M.~M.}\ \bibnamefont
  {Wilde}},\ }\href@noop {} {\emph {\bibinfo {title} {Quantum Information
  Theory}}}\ (\bibinfo  {publisher} {Cambridge University Press},\ \bibinfo
  {year} {2017})\BibitemShut {NoStop}%
\bibitem [{\citenamefont {Holevo}(2013)}]{holevo2013quantum}%
  \BibitemOpen
  \bibfield  {author} {\bibinfo {author} {\bibfnamefont {A.~S.}\ \bibnamefont
  {Holevo}},\ }\href@noop {} {\emph {\bibinfo {title} {Quantum Systems,
  Channels, Information: A Mathematical Introduction}}},\ De Gruyter Studies in
  Mathematical Physics\ (\bibinfo  {publisher} {De Gruyter},\ \bibinfo {year}
  {2013})\BibitemShut {NoStop}%
\bibitem [{\citenamefont {Tomamichel}(2015)}]{tomamichel2015quantum}%
  \BibitemOpen
  \bibfield  {author} {\bibinfo {author} {\bibfnamefont {M.}~\bibnamefont
  {Tomamichel}},\ }\href@noop {} {\emph {\bibinfo {title} {Quantum Information
  Processing with Finite Resources: Mathematical Foundations}}},\
  SpringerBriefs in Mathematical Physics\ (\bibinfo  {publisher} {Springer
  International Publishing},\ \bibinfo {year} {2015})\BibitemShut {NoStop}%
\bibitem [{\citenamefont {Audenaert}\ \emph {et~al.}(2001)\citenamefont
  {Audenaert}, \citenamefont {Eisert}, \citenamefont {Jan{\'e}}, \citenamefont
  {Plenio}, \citenamefont {Virmani},\ and\ \citenamefont
  {De~Moor}}]{audenaert2001asymptotic}%
  \BibitemOpen
  \bibfield  {author} {\bibinfo {author} {\bibfnamefont {K.}~\bibnamefont
  {Audenaert}}, \bibinfo {author} {\bibfnamefont {J.}~\bibnamefont {Eisert}},
  \bibinfo {author} {\bibfnamefont {E.}~\bibnamefont {Jan{\'e}}}, \bibinfo
  {author} {\bibfnamefont {M.~B.}\ \bibnamefont {Plenio}}, \bibinfo {author}
  {\bibfnamefont {S.}~\bibnamefont {Virmani}}, \ and\ \bibinfo {author}
  {\bibfnamefont {B.}~\bibnamefont {De~Moor}},\ }\href@noop {} {\bibfield
  {journal} {\bibinfo  {journal} {Physical review letters}\ }\textbf {\bibinfo
  {volume} {87}},\ \bibinfo {pages} {217902} (\bibinfo {year}
  {2001})}\BibitemShut {NoStop}%
\bibitem [{\citenamefont {Vedral}(2002)}]{vedral2002role}%
  \BibitemOpen
  \bibfield  {author} {\bibinfo {author} {\bibfnamefont {V.}~\bibnamefont
  {Vedral}},\ }\href@noop {} {\bibfield  {journal} {\bibinfo  {journal}
  {Reviews of Modern Physics}\ }\textbf {\bibinfo {volume} {74}},\ \bibinfo
  {pages} {197} (\bibinfo {year} {2002})}\BibitemShut {NoStop}%
\bibitem [{\citenamefont {Christandl}\ \emph {et~al.}(2007)\citenamefont
  {Christandl}, \citenamefont {Ekert}, \citenamefont {Horodecki}, \citenamefont
  {Horodecki}, \citenamefont {Oppenheim},\ and\ \citenamefont
  {Renner}}]{christandl2007unifying}%
  \BibitemOpen
  \bibfield  {author} {\bibinfo {author} {\bibfnamefont {M.}~\bibnamefont
  {Christandl}}, \bibinfo {author} {\bibfnamefont {A.}~\bibnamefont {Ekert}},
  \bibinfo {author} {\bibfnamefont {M.}~\bibnamefont {Horodecki}}, \bibinfo
  {author} {\bibfnamefont {P.}~\bibnamefont {Horodecki}}, \bibinfo {author}
  {\bibfnamefont {J.}~\bibnamefont {Oppenheim}}, \ and\ \bibinfo {author}
  {\bibfnamefont {R.}~\bibnamefont {Renner}},\ }in\ \href@noop {} {\emph
  {\bibinfo {booktitle} {Theory of Cryptography Conference}}}\ (\bibinfo
  {organization} {Springer},\ \bibinfo {year} {2007})\ pp.\ \bibinfo {pages}
  {456--478}\BibitemShut {NoStop}%
\bibitem [{\citenamefont {Li}\ and\ \citenamefont
  {Winter}(2014)}]{li2014relative}%
  \BibitemOpen
  \bibfield  {author} {\bibinfo {author} {\bibfnamefont {K.}~\bibnamefont
  {Li}}\ and\ \bibinfo {author} {\bibfnamefont {A.}~\bibnamefont {Winter}},\
  }\href@noop {} {\bibfield  {journal} {\bibinfo  {journal} {Communications in
  Mathematical Physics}\ }\textbf {\bibinfo {volume} {326}},\ \bibinfo {pages}
  {63} (\bibinfo {year} {2014})}\BibitemShut {NoStop}%
\bibitem [{\citenamefont {Wilde}(2016)}]{wilde2016squashed}%
  \BibitemOpen
  \bibfield  {author} {\bibinfo {author} {\bibfnamefont {M.~M.}\ \bibnamefont
  {Wilde}},\ }\href@noop {} {\bibfield  {journal} {\bibinfo  {journal} {Quantum
  Information Processing}\ }\textbf {\bibinfo {volume} {15}},\ \bibinfo {pages}
  {4563} (\bibinfo {year} {2016})}\BibitemShut {NoStop}%
\bibitem [{\citenamefont {Seshadreesan}\ and\ \citenamefont
  {Wilde}(2015)}]{seshadreesan2015fidelity}%
  \BibitemOpen
  \bibfield  {author} {\bibinfo {author} {\bibfnamefont {K.~P.}\ \bibnamefont
  {Seshadreesan}}\ and\ \bibinfo {author} {\bibfnamefont {M.~M.}\ \bibnamefont
  {Wilde}},\ }\href@noop {} {\bibfield  {journal} {\bibinfo  {journal}
  {Physical Review A}\ }\textbf {\bibinfo {volume} {92}},\ \bibinfo {pages}
  {042321} (\bibinfo {year} {2015})}\BibitemShut {NoStop}%
\bibitem [{\citenamefont {Li}\ and\ \citenamefont
  {Winter}(2018)}]{li2018squashed}%
  \BibitemOpen
  \bibfield  {author} {\bibinfo {author} {\bibfnamefont {K.}~\bibnamefont
  {Li}}\ and\ \bibinfo {author} {\bibfnamefont {A.}~\bibnamefont {Winter}},\
  }\href@noop {} {\bibfield  {journal} {\bibinfo  {journal} {Foundations of
  Physics}\ }\textbf {\bibinfo {volume} {48}},\ \bibinfo {pages} {910}
  (\bibinfo {year} {2018})}\BibitemShut {NoStop}%
\bibitem [{\citenamefont {Adesso}\ \emph {et~al.}(2007)\citenamefont {Adesso},
  \citenamefont {Ericsson},\ and\ \citenamefont
  {Illuminati}}]{adesso2007coexistence}%
  \BibitemOpen
  \bibfield  {author} {\bibinfo {author} {\bibfnamefont {G.}~\bibnamefont
  {Adesso}}, \bibinfo {author} {\bibfnamefont {M.}~\bibnamefont {Ericsson}}, \
  and\ \bibinfo {author} {\bibfnamefont {F.}~\bibnamefont {Illuminati}},\
  }\href@noop {} {\bibfield  {journal} {\bibinfo  {journal} {Physical Review
  A}\ }\textbf {\bibinfo {volume} {76}},\ \bibinfo {pages} {022315} (\bibinfo
  {year} {2007})}\BibitemShut {NoStop}%
\bibitem [{\citenamefont {Avis}\ \emph {et~al.}(2008)\citenamefont {Avis},
  \citenamefont {Hayden},\ and\ \citenamefont {Savov}}]{avis2008distributed}%
  \BibitemOpen
  \bibfield  {author} {\bibinfo {author} {\bibfnamefont {D.}~\bibnamefont
  {Avis}}, \bibinfo {author} {\bibfnamefont {P.}~\bibnamefont {Hayden}}, \ and\
  \bibinfo {author} {\bibfnamefont {I.}~\bibnamefont {Savov}},\ }\href@noop {}
  {\bibfield  {journal} {\bibinfo  {journal} {Journal of Physics A:
  Mathematical and Theoretical}\ }\textbf {\bibinfo {volume} {41}},\ \bibinfo
  {pages} {115301} (\bibinfo {year} {2008})}\BibitemShut {NoStop}%
\bibitem [{\citenamefont {Yang}\ \emph {et~al.}(2009)\citenamefont {Yang},
  \citenamefont {Horodecki}, \citenamefont {Horodecki}, \citenamefont
  {Horodecki}, \citenamefont {Oppenheim},\ and\ \citenamefont
  {Song}}]{yang2009squashed}%
  \BibitemOpen
  \bibfield  {author} {\bibinfo {author} {\bibfnamefont {D.}~\bibnamefont
  {Yang}}, \bibinfo {author} {\bibfnamefont {K.}~\bibnamefont {Horodecki}},
  \bibinfo {author} {\bibfnamefont {M.}~\bibnamefont {Horodecki}}, \bibinfo
  {author} {\bibfnamefont {P.}~\bibnamefont {Horodecki}}, \bibinfo {author}
  {\bibfnamefont {J.}~\bibnamefont {Oppenheim}}, \ and\ \bibinfo {author}
  {\bibfnamefont {W.}~\bibnamefont {Song}},\ }\href@noop {} {\bibfield
  {journal} {\bibinfo  {journal} {IEEE Transactions on Information Theory}\
  }\textbf {\bibinfo {volume} {55}},\ \bibinfo {pages} {3375} (\bibinfo {year}
  {2009})}\BibitemShut {NoStop}%
\bibitem [{\citenamefont {Pirandola}\ \emph {et~al.}(2017)\citenamefont
  {Pirandola}, \citenamefont {Laurenza}, \citenamefont {Ottaviani},\ and\
  \citenamefont {Banchi}}]{pirandola2017fundamental}%
  \BibitemOpen
  \bibfield  {author} {\bibinfo {author} {\bibfnamefont {S.}~\bibnamefont
  {Pirandola}}, \bibinfo {author} {\bibfnamefont {R.}~\bibnamefont {Laurenza}},
  \bibinfo {author} {\bibfnamefont {C.}~\bibnamefont {Ottaviani}}, \ and\
  \bibinfo {author} {\bibfnamefont {L.}~\bibnamefont {Banchi}},\ }\href@noop {}
  {\bibfield  {journal} {\bibinfo  {journal} {Nature communications}\ }\textbf
  {\bibinfo {volume} {8}},\ \bibinfo {pages} {15043} (\bibinfo {year}
  {2017})}\BibitemShut {NoStop}%
\bibitem [{\citenamefont {Takeoka}\ \emph
  {et~al.}(2014{\natexlab{a}})\citenamefont {Takeoka}, \citenamefont {Guha},\
  and\ \citenamefont {Wilde}}]{takeoka2014squashed}%
  \BibitemOpen
  \bibfield  {author} {\bibinfo {author} {\bibfnamefont {M.}~\bibnamefont
  {Takeoka}}, \bibinfo {author} {\bibfnamefont {S.}~\bibnamefont {Guha}}, \
  and\ \bibinfo {author} {\bibfnamefont {M.~M.}\ \bibnamefont {Wilde}},\
  }\href@noop {} {\bibfield  {journal} {\bibinfo  {journal} {IEEE Transactions
  on Information Theory}\ }\textbf {\bibinfo {volume} {60}},\ \bibinfo {pages}
  {4987} (\bibinfo {year} {2014}{\natexlab{a}})}\BibitemShut {NoStop}%
\bibitem [{\citenamefont {Berta}\ and\ \citenamefont
  {Wilde}(2018)}]{berta2018amortization}%
  \BibitemOpen
  \bibfield  {author} {\bibinfo {author} {\bibfnamefont {M.}~\bibnamefont
  {Berta}}\ and\ \bibinfo {author} {\bibfnamefont {M.~M.}\ \bibnamefont
  {Wilde}},\ }\href@noop {} {\bibfield  {journal} {\bibinfo  {journal} {New
  Journal of Physics}\ }\textbf {\bibinfo {volume} {20}},\ \bibinfo {pages}
  {053044} (\bibinfo {year} {2018})}\BibitemShut {NoStop}%
\bibitem [{\citenamefont {Lami}\ \emph {et~al.}(2017)\citenamefont {Lami},
  \citenamefont {Hirche}, \citenamefont {Adesso},\ and\ \citenamefont
  {Winter}}]{lami2017log}%
  \BibitemOpen
  \bibfield  {author} {\bibinfo {author} {\bibfnamefont {L.}~\bibnamefont
  {Lami}}, \bibinfo {author} {\bibfnamefont {C.}~\bibnamefont {Hirche}},
  \bibinfo {author} {\bibfnamefont {G.}~\bibnamefont {Adesso}}, \ and\ \bibinfo
  {author} {\bibfnamefont {A.}~\bibnamefont {Winter}},\ }\href@noop {}
  {\bibfield  {journal} {\bibinfo  {journal} {IEEE Transactions on Information
  Theory}\ }\textbf {\bibinfo {volume} {63}},\ \bibinfo {pages} {7553}
  (\bibinfo {year} {2017})}\BibitemShut {NoStop}%
\bibitem [{\citenamefont {K{\"o}nig}(2015)}]{koenig2015conditional}%
  \BibitemOpen
  \bibfield  {author} {\bibinfo {author} {\bibfnamefont {R.}~\bibnamefont
  {K{\"o}nig}},\ }\href@noop {} {\bibfield  {journal} {\bibinfo  {journal}
  {Journal of Mathematical Physics}\ }\textbf {\bibinfo {volume} {56}},\
  \bibinfo {pages} {022201} (\bibinfo {year} {2015})}\BibitemShut {NoStop}%
\bibitem [{\citenamefont {De~Palma}\ and\ \citenamefont
  {Trevisan}(2018)}]{de2018conditional}%
  \BibitemOpen
  \bibfield  {author} {\bibinfo {author} {\bibfnamefont {G.}~\bibnamefont
  {De~Palma}}\ and\ \bibinfo {author} {\bibfnamefont {D.}~\bibnamefont
  {Trevisan}},\ }\href
  {https://link.springer.com/article/10.1007\%2Fs00220-017-3082-8} {\bibfield
  {journal} {\bibinfo  {journal} {Communications in Mathematical Physics}\
  }\textbf {\bibinfo {volume} {360}},\ \bibinfo {pages} {639} (\bibinfo {year}
  {2018})}\BibitemShut {NoStop}%
\bibitem [{\citenamefont {De~Palma}\ and\ \citenamefont
  {Huber}(2018)}]{huber2018conditional}%
  \BibitemOpen
  \bibfield  {author} {\bibinfo {author} {\bibfnamefont {G.}~\bibnamefont
  {De~Palma}}\ and\ \bibinfo {author} {\bibfnamefont {S.}~\bibnamefont
  {Huber}},\ }\href {https://aip.scitation.org/doi/10.1063/1.5027495}
  {\bibfield  {journal} {\bibinfo  {journal} {Journal of Mathematical Physics}\
  }\textbf {\bibinfo {volume} {59}},\ \bibinfo {pages} {122201} (\bibinfo
  {year} {2018})}\BibitemShut {NoStop}%
\bibitem [{\citenamefont {De~Palma}(2019)}]{de2019entropy}%
  \BibitemOpen
  \bibfield  {author} {\bibinfo {author} {\bibfnamefont {G.}~\bibnamefont
  {De~Palma}},\ }\href
  {https://iopscience.iop.org/article/10.1088/1751-8121/aafff4} {\bibfield
  {journal} {\bibinfo  {journal} {Journal of Physics A: Mathematical and
  Theoretical}\ } (\bibinfo {year} {2019})}\BibitemShut {NoStop}%
\bibitem [{\citenamefont {Takeoka}\ \emph
  {et~al.}(2014{\natexlab{b}})\citenamefont {Takeoka}, \citenamefont {Guha},\
  and\ \citenamefont {Wilde}}]{takeoka2014fundamental}%
  \BibitemOpen
  \bibfield  {author} {\bibinfo {author} {\bibfnamefont {M.}~\bibnamefont
  {Takeoka}}, \bibinfo {author} {\bibfnamefont {S.}~\bibnamefont {Guha}}, \
  and\ \bibinfo {author} {\bibfnamefont {M.~M.}\ \bibnamefont {Wilde}},\
  }\href@noop {} {\bibfield  {journal} {\bibinfo  {journal} {Nature
  communications}\ }\textbf {\bibinfo {volume} {5}},\ \bibinfo {pages} {5235}
  (\bibinfo {year} {2014}{\natexlab{b}})}\BibitemShut {NoStop}%
\bibitem [{\citenamefont {Goodenough}\ \emph {et~al.}(2016)\citenamefont
  {Goodenough}, \citenamefont {Elkouss},\ and\ \citenamefont
  {Wehner}}]{goodenough2016assessing}%
  \BibitemOpen
  \bibfield  {author} {\bibinfo {author} {\bibfnamefont {K.}~\bibnamefont
  {Goodenough}}, \bibinfo {author} {\bibfnamefont {D.}~\bibnamefont {Elkouss}},
  \ and\ \bibinfo {author} {\bibfnamefont {S.}~\bibnamefont {Wehner}},\
  }\href@noop {} {\bibfield  {journal} {\bibinfo  {journal} {New Journal of
  Physics}\ }\textbf {\bibinfo {volume} {18}},\ \bibinfo {pages} {063005}
  (\bibinfo {year} {2016})}\BibitemShut {NoStop}%
\bibitem [{\citenamefont {Davis}\ \emph {et~al.}(2018)\citenamefont {Davis},
  \citenamefont {Shirokov},\ and\ \citenamefont {Wilde}}]{davis2018energy}%
  \BibitemOpen
  \bibfield  {author} {\bibinfo {author} {\bibfnamefont {N.}~\bibnamefont
  {Davis}}, \bibinfo {author} {\bibfnamefont {M.~E.}\ \bibnamefont {Shirokov}},
  \ and\ \bibinfo {author} {\bibfnamefont {M.~M.}\ \bibnamefont {Wilde}},\
  }\href@noop {} {\bibfield  {journal} {\bibinfo  {journal} {Physical Review
  A}\ }\textbf {\bibinfo {volume} {97}},\ \bibinfo {pages} {062310} (\bibinfo
  {year} {2018})}\BibitemShut {NoStop}%
\bibitem [{\citenamefont {Caruso}\ \emph {et~al.}(2006)\citenamefont {Caruso},
  \citenamefont {Giovannetti},\ and\ \citenamefont {Holevo}}]{caruso2006one}%
  \BibitemOpen
  \bibfield  {author} {\bibinfo {author} {\bibfnamefont {F.}~\bibnamefont
  {Caruso}}, \bibinfo {author} {\bibfnamefont {V.}~\bibnamefont {Giovannetti}},
  \ and\ \bibinfo {author} {\bibfnamefont {A.~S.}\ \bibnamefont {Holevo}},\
  }\href@noop {} {\bibfield  {journal} {\bibinfo  {journal} {New Journal of
  Physics}\ }\textbf {\bibinfo {volume} {8}},\ \bibinfo {pages} {310} (\bibinfo
  {year} {2006})}\BibitemShut {NoStop}%
\bibitem [{\citenamefont {Caruso}\ and\ \citenamefont
  {Giovannetti}(2006)}]{caruso2006degradability}%
  \BibitemOpen
  \bibfield  {author} {\bibinfo {author} {\bibfnamefont {F.}~\bibnamefont
  {Caruso}}\ and\ \bibinfo {author} {\bibfnamefont {V.}~\bibnamefont
  {Giovannetti}},\ }\href@noop {} {\bibfield  {journal} {\bibinfo  {journal}
  {Physical Review A}\ }\textbf {\bibinfo {volume} {74}},\ \bibinfo {pages}
  {062307} (\bibinfo {year} {2006})}\BibitemShut {NoStop}%
\bibitem [{\citenamefont {Holevo}(2007)}]{holevo2007one}%
  \BibitemOpen
  \bibfield  {author} {\bibinfo {author} {\bibfnamefont {A.~S.}\ \bibnamefont
  {Holevo}},\ }\href@noop {} {\bibfield  {journal} {\bibinfo  {journal}
  {Problems of Information Transmission}\ }\textbf {\bibinfo {volume} {43}},\
  \bibinfo {pages} {1} (\bibinfo {year} {2007})}\BibitemShut {NoStop}%
\bibitem [{\citenamefont {Weedbrook}\ \emph {et~al.}(2012)\citenamefont
  {Weedbrook}, \citenamefont {Pirandola}, \citenamefont {Garcia-Patron},
  \citenamefont {Cerf}, \citenamefont {Ralph}, \citenamefont {Shapiro},\ and\
  \citenamefont {Lloyd}}]{weedbrook2012gaussian}%
  \BibitemOpen
  \bibfield  {author} {\bibinfo {author} {\bibfnamefont {C.}~\bibnamefont
  {Weedbrook}}, \bibinfo {author} {\bibfnamefont {S.}~\bibnamefont
  {Pirandola}}, \bibinfo {author} {\bibfnamefont {R.}~\bibnamefont
  {Garcia-Patron}}, \bibinfo {author} {\bibfnamefont {N.~J.}\ \bibnamefont
  {Cerf}}, \bibinfo {author} {\bibfnamefont {T.~C.}\ \bibnamefont {Ralph}},
  \bibinfo {author} {\bibfnamefont {J.~H.}\ \bibnamefont {Shapiro}}, \ and\
  \bibinfo {author} {\bibfnamefont {S.}~\bibnamefont {Lloyd}},\ }\href@noop {}
  {\bibfield  {journal} {\bibinfo  {journal} {Reviews of Modern Physics}\
  }\textbf {\bibinfo {volume} {84}},\ \bibinfo {pages} {621} (\bibinfo {year}
  {2012})}\BibitemShut {NoStop}%
\bibitem [{\citenamefont {Serafini}(2017)}]{serafini2017quantum}%
  \BibitemOpen
  \bibfield  {author} {\bibinfo {author} {\bibfnamefont {A.}~\bibnamefont
  {Serafini}},\ }\href@noop {} {\emph {\bibinfo {title} {Quantum Continuous
  Variables: A Primer of Theoretical Methods}}}\ (\bibinfo  {publisher} {CRC
  Press},\ \bibinfo {year} {2017})\BibitemShut {NoStop}%
\bibitem [{\citenamefont {Bennett}\ and\ \citenamefont
  {Brassard}(1984)}]{bennett1984proceedings}%
  \BibitemOpen
  \bibfield  {author} {\bibinfo {author} {\bibfnamefont {C.~H.}\ \bibnamefont
  {Bennett}}\ and\ \bibinfo {author} {\bibfnamefont {G.}~\bibnamefont
  {Brassard}},\ }\href@noop {} {\enquote {\bibinfo {title} {Proceedings of the
  ieee international conference on computers, systems and signal processing},}\
  } (\bibinfo {year} {1984})\BibitemShut {NoStop}%
\bibitem [{\citenamefont {Ekert}(1991)}]{ekert1991quantum}%
  \BibitemOpen
  \bibfield  {author} {\bibinfo {author} {\bibfnamefont {A.~K.}\ \bibnamefont
  {Ekert}},\ }\href@noop {} {\bibfield  {journal} {\bibinfo  {journal}
  {Physical review letters}\ }\textbf {\bibinfo {volume} {67}},\ \bibinfo
  {pages} {661} (\bibinfo {year} {1991})}\BibitemShut {NoStop}%
\bibitem [{\citenamefont {Gisin}\ \emph {et~al.}(2002)\citenamefont {Gisin},
  \citenamefont {Ribordy}, \citenamefont {Tittel},\ and\ \citenamefont
  {Zbinden}}]{gisin2002quantum}%
  \BibitemOpen
  \bibfield  {author} {\bibinfo {author} {\bibfnamefont {N.}~\bibnamefont
  {Gisin}}, \bibinfo {author} {\bibfnamefont {G.}~\bibnamefont {Ribordy}},
  \bibinfo {author} {\bibfnamefont {W.}~\bibnamefont {Tittel}}, \ and\ \bibinfo
  {author} {\bibfnamefont {H.}~\bibnamefont {Zbinden}},\ }\href@noop {}
  {\bibfield  {journal} {\bibinfo  {journal} {Reviews of modern physics}\
  }\textbf {\bibinfo {volume} {74}},\ \bibinfo {pages} {145} (\bibinfo {year}
  {2002})}\BibitemShut {NoStop}%
\bibitem [{\citenamefont {Lloyd}\ \emph {et~al.}(2004)\citenamefont {Lloyd},
  \citenamefont {Shapiro}, \citenamefont {Wong}, \citenamefont {Kumar},
  \citenamefont {Shahriar},\ and\ \citenamefont
  {Yuen}}]{lloyd2004infrastructure}%
  \BibitemOpen
  \bibfield  {author} {\bibinfo {author} {\bibfnamefont {S.}~\bibnamefont
  {Lloyd}}, \bibinfo {author} {\bibfnamefont {J.~H.}\ \bibnamefont {Shapiro}},
  \bibinfo {author} {\bibfnamefont {F.~N.}\ \bibnamefont {Wong}}, \bibinfo
  {author} {\bibfnamefont {P.}~\bibnamefont {Kumar}}, \bibinfo {author}
  {\bibfnamefont {S.~M.}\ \bibnamefont {Shahriar}}, \ and\ \bibinfo {author}
  {\bibfnamefont {H.~P.}\ \bibnamefont {Yuen}},\ }\href@noop {} {\bibfield
  {journal} {\bibinfo  {journal} {ACM SIGCOMM Computer Communication Review}\
  }\textbf {\bibinfo {volume} {34}},\ \bibinfo {pages} {9} (\bibinfo {year}
  {2004})}\BibitemShut {NoStop}%
\bibitem [{\citenamefont {Braunstein}\ and\ \citenamefont
  {Van~Loock}(2005)}]{braunstein2005quantum}%
  \BibitemOpen
  \bibfield  {author} {\bibinfo {author} {\bibfnamefont {S.~L.}\ \bibnamefont
  {Braunstein}}\ and\ \bibinfo {author} {\bibfnamefont {P.}~\bibnamefont
  {Van~Loock}},\ }\href@noop {} {\bibfield  {journal} {\bibinfo  {journal}
  {Reviews of Modern Physics}\ }\textbf {\bibinfo {volume} {77}},\ \bibinfo
  {pages} {513} (\bibinfo {year} {2005})}\BibitemShut {NoStop}%
\bibitem [{\citenamefont {Kimble}(2008)}]{kimble2008quantum}%
  \BibitemOpen
  \bibfield  {author} {\bibinfo {author} {\bibfnamefont {H.~J.}\ \bibnamefont
  {Kimble}},\ }\href@noop {} {\bibfield  {journal} {\bibinfo  {journal}
  {Nature}\ }\textbf {\bibinfo {volume} {453}},\ \bibinfo {pages} {1023}
  (\bibinfo {year} {2008})}\BibitemShut {NoStop}%
\bibitem [{\citenamefont {Scarani}\ \emph {et~al.}(2009)\citenamefont
  {Scarani}, \citenamefont {Bechmann-Pasquinucci}, \citenamefont {Cerf},
  \citenamefont {Du{\v{s}}ek}, \citenamefont {L{\"u}tkenhaus},\ and\
  \citenamefont {Peev}}]{scarani2009security}%
  \BibitemOpen
  \bibfield  {author} {\bibinfo {author} {\bibfnamefont {V.}~\bibnamefont
  {Scarani}}, \bibinfo {author} {\bibfnamefont {H.}~\bibnamefont
  {Bechmann-Pasquinucci}}, \bibinfo {author} {\bibfnamefont {N.~J.}\
  \bibnamefont {Cerf}}, \bibinfo {author} {\bibfnamefont {M.}~\bibnamefont
  {Du{\v{s}}ek}}, \bibinfo {author} {\bibfnamefont {N.}~\bibnamefont
  {L{\"u}tkenhaus}}, \ and\ \bibinfo {author} {\bibfnamefont {M.}~\bibnamefont
  {Peev}},\ }\href@noop {} {\bibfield  {journal} {\bibinfo  {journal} {Reviews
  of modern physics}\ }\textbf {\bibinfo {volume} {81}},\ \bibinfo {pages}
  {1301} (\bibinfo {year} {2009})}\BibitemShut {NoStop}%
\bibitem [{\citenamefont {Pirandola}\ and\ \citenamefont
  {Braunstein}(2016)}]{pirandola2016physics}%
  \BibitemOpen
  \bibfield  {author} {\bibinfo {author} {\bibfnamefont {S.}~\bibnamefont
  {Pirandola}}\ and\ \bibinfo {author} {\bibfnamefont {S.~L.}\ \bibnamefont
  {Braunstein}},\ }\href@noop {} {\bibfield  {journal} {\bibinfo  {journal}
  {Nature News}\ }\textbf {\bibinfo {volume} {532}},\ \bibinfo {pages} {169}
  (\bibinfo {year} {2016})}\BibitemShut {NoStop}%
\bibitem [{\citenamefont {Pirandola}(2016)}]{pirandola2016capacities}%
  \BibitemOpen
  \bibfield  {author} {\bibinfo {author} {\bibfnamefont {S.}~\bibnamefont
  {Pirandola}},\ }\href@noop {} {\bibfield  {journal} {\bibinfo  {journal}
  {arXiv preprint arXiv:1601.00966}\ } (\bibinfo {year} {2016})}\BibitemShut
  {NoStop}%
\bibitem [{\citenamefont {Azuma}\ \emph {et~al.}(2016)\citenamefont {Azuma},
  \citenamefont {Mizutani},\ and\ \citenamefont {Lo}}]{azuma2016fundamental}%
  \BibitemOpen
  \bibfield  {author} {\bibinfo {author} {\bibfnamefont {K.}~\bibnamefont
  {Azuma}}, \bibinfo {author} {\bibfnamefont {A.}~\bibnamefont {Mizutani}}, \
  and\ \bibinfo {author} {\bibfnamefont {H.-K.}\ \bibnamefont {Lo}},\
  }\href@noop {} {\bibfield  {journal} {\bibinfo  {journal} {Nature
  communications}\ }\textbf {\bibinfo {volume} {7}},\ \bibinfo {pages} {13523}
  (\bibinfo {year} {2016})}\BibitemShut {NoStop}%
\bibitem [{\citenamefont {Laurenza}\ and\ \citenamefont
  {Pirandola}(2017)}]{laurenza2017general}%
  \BibitemOpen
  \bibfield  {author} {\bibinfo {author} {\bibfnamefont {R.}~\bibnamefont
  {Laurenza}}\ and\ \bibinfo {author} {\bibfnamefont {S.}~\bibnamefont
  {Pirandola}},\ }\href@noop {} {\bibfield  {journal} {\bibinfo  {journal}
  {Physical Review A}\ }\textbf {\bibinfo {volume} {96}},\ \bibinfo {pages}
  {032318} (\bibinfo {year} {2017})}\BibitemShut {NoStop}%
\bibitem [{\citenamefont {Laurenza}\ \emph {et~al.}(2018)\citenamefont
  {Laurenza}, \citenamefont {Braunstein},\ and\ \citenamefont
  {Pirandola}}]{laurenza2018finite}%
  \BibitemOpen
  \bibfield  {author} {\bibinfo {author} {\bibfnamefont {R.}~\bibnamefont
  {Laurenza}}, \bibinfo {author} {\bibfnamefont {S.~L.}\ \bibnamefont
  {Braunstein}}, \ and\ \bibinfo {author} {\bibfnamefont {S.}~\bibnamefont
  {Pirandola}},\ }\href@noop {} {\bibfield  {journal} {\bibinfo  {journal}
  {Scientific reports}\ }\textbf {\bibinfo {volume} {8}},\ \bibinfo {pages}
  {15267} (\bibinfo {year} {2018})}\BibitemShut {NoStop}%
\bibitem [{\citenamefont {Cope}\ \emph {et~al.}(2018)\citenamefont {Cope},
  \citenamefont {Goodenough},\ and\ \citenamefont
  {Pirandola}}]{cope2018converse}%
  \BibitemOpen
  \bibfield  {author} {\bibinfo {author} {\bibfnamefont {T.~P.}\ \bibnamefont
  {Cope}}, \bibinfo {author} {\bibfnamefont {K.}~\bibnamefont {Goodenough}}, \
  and\ \bibinfo {author} {\bibfnamefont {S.}~\bibnamefont {Pirandola}},\
  }\href@noop {} {\bibfield  {journal} {\bibinfo  {journal} {Journal of Physics
  A: Mathematical and Theoretical}\ }\textbf {\bibinfo {volume} {51}},\
  \bibinfo {pages} {494001} (\bibinfo {year} {2018})}\BibitemShut {NoStop}%
\bibitem [{\citenamefont {Wehner}\ \emph {et~al.}(2018)\citenamefont {Wehner},
  \citenamefont {Elkouss},\ and\ \citenamefont {Hanson}}]{wehner2018quantum}%
  \BibitemOpen
  \bibfield  {author} {\bibinfo {author} {\bibfnamefont {S.}~\bibnamefont
  {Wehner}}, \bibinfo {author} {\bibfnamefont {D.}~\bibnamefont {Elkouss}}, \
  and\ \bibinfo {author} {\bibfnamefont {R.}~\bibnamefont {Hanson}},\
  }\href@noop {} {\bibfield  {journal} {\bibinfo  {journal} {Science}\ }\textbf
  {\bibinfo {volume} {362}},\ \bibinfo {pages} {eaam9288} (\bibinfo {year}
  {2018})}\BibitemShut {NoStop}%
\bibitem [{\citenamefont {Laurenza}\ \emph {et~al.}(2019)\citenamefont
  {Laurenza}, \citenamefont {Tserkis}, \citenamefont {Banchi}, \citenamefont
  {Braunstein}, \citenamefont {Ralph},\ and\ \citenamefont
  {Pirandola}}]{laurenza2019tight}%
  \BibitemOpen
  \bibfield  {author} {\bibinfo {author} {\bibfnamefont {R.}~\bibnamefont
  {Laurenza}}, \bibinfo {author} {\bibfnamefont {S.}~\bibnamefont {Tserkis}},
  \bibinfo {author} {\bibfnamefont {L.}~\bibnamefont {Banchi}}, \bibinfo
  {author} {\bibfnamefont {S.~L.}\ \bibnamefont {Braunstein}}, \bibinfo
  {author} {\bibfnamefont {T.~C.}\ \bibnamefont {Ralph}}, \ and\ \bibinfo
  {author} {\bibfnamefont {S.}~\bibnamefont {Pirandola}},\ }\href@noop {}
  {\bibfield  {journal} {\bibinfo  {journal} {Physical Review A}\ }\textbf
  {\bibinfo {volume} {100}},\ \bibinfo {pages} {042301} (\bibinfo {year}
  {2019})}\BibitemShut {NoStop}%
\bibitem [{\citenamefont {Pirandola}(2019)}]{pirandola2019end}%
  \BibitemOpen
  \bibfield  {author} {\bibinfo {author} {\bibfnamefont {S.}~\bibnamefont
  {Pirandola}},\ }\href@noop {} {\bibfield  {journal} {\bibinfo  {journal}
  {Communications Physics}\ }\textbf {\bibinfo {volume} {2}},\ \bibinfo {pages}
  {51} (\bibinfo {year} {2019})}\BibitemShut {NoStop}%
\bibitem [{\citenamefont {Pirandola}\ \emph {et~al.}(2019)\citenamefont
  {Pirandola}, \citenamefont {Andersen}, \citenamefont {Banchi}, \citenamefont
  {Berta}, \citenamefont {Bunandar}, \citenamefont {Colbeck}, \citenamefont
  {Englund}, \citenamefont {Gehring}, \citenamefont {Lupo}, \citenamefont
  {Ottaviani} \emph {et~al.}}]{pirandola2019advances}%
  \BibitemOpen
  \bibfield  {author} {\bibinfo {author} {\bibfnamefont {S.}~\bibnamefont
  {Pirandola}}, \bibinfo {author} {\bibfnamefont {U.}~\bibnamefont {Andersen}},
  \bibinfo {author} {\bibfnamefont {L.}~\bibnamefont {Banchi}}, \bibinfo
  {author} {\bibfnamefont {M.}~\bibnamefont {Berta}}, \bibinfo {author}
  {\bibfnamefont {D.}~\bibnamefont {Bunandar}}, \bibinfo {author}
  {\bibfnamefont {R.}~\bibnamefont {Colbeck}}, \bibinfo {author} {\bibfnamefont
  {D.}~\bibnamefont {Englund}}, \bibinfo {author} {\bibfnamefont
  {T.}~\bibnamefont {Gehring}}, \bibinfo {author} {\bibfnamefont
  {C.}~\bibnamefont {Lupo}}, \bibinfo {author} {\bibfnamefont {C.}~\bibnamefont
  {Ottaviani}},  \emph {et~al.},\ }\href@noop {} {\bibfield  {journal}
  {\bibinfo  {journal} {arXiv preprint arXiv:1906.01645}\ } (\bibinfo {year}
  {2019})}\BibitemShut {NoStop}%
\bibitem [{\citenamefont {Pirandola}\ \emph {et~al.}(2009)\citenamefont
  {Pirandola}, \citenamefont {Garc{\'\i}a-Patr{\'o}n}, \citenamefont
  {Braunstein},\ and\ \citenamefont {Lloyd}}]{pirandola2009direct}%
  \BibitemOpen
  \bibfield  {author} {\bibinfo {author} {\bibfnamefont {S.}~\bibnamefont
  {Pirandola}}, \bibinfo {author} {\bibfnamefont {R.}~\bibnamefont
  {Garc{\'\i}a-Patr{\'o}n}}, \bibinfo {author} {\bibfnamefont {S.~L.}\
  \bibnamefont {Braunstein}}, \ and\ \bibinfo {author} {\bibfnamefont
  {S.}~\bibnamefont {Lloyd}},\ }\href@noop {} {\bibfield  {journal} {\bibinfo
  {journal} {Physical review letters}\ }\textbf {\bibinfo {volume} {102}},\
  \bibinfo {pages} {050503} (\bibinfo {year} {2009})}\BibitemShut {NoStop}%
\bibitem [{\citenamefont {Wilde}\ \emph {et~al.}(2017)\citenamefont {Wilde},
  \citenamefont {Tomamichel},\ and\ \citenamefont {Berta}}]{wilde2017converse}%
  \BibitemOpen
  \bibfield  {author} {\bibinfo {author} {\bibfnamefont {M.~M.}\ \bibnamefont
  {Wilde}}, \bibinfo {author} {\bibfnamefont {M.}~\bibnamefont {Tomamichel}}, \
  and\ \bibinfo {author} {\bibfnamefont {M.}~\bibnamefont {Berta}},\
  }\href@noop {} {\bibfield  {journal} {\bibinfo  {journal} {IEEE Transactions
  on Information Theory}\ }\textbf {\bibinfo {volume} {63}},\ \bibinfo {pages}
  {1792} (\bibinfo {year} {2017})}\BibitemShut {NoStop}%
\bibitem [{\citenamefont {Pirandola}\ \emph {et~al.}(2018)\citenamefont
  {Pirandola}, \citenamefont {Braunstein}, \citenamefont {Laurenza},
  \citenamefont {Ottaviani}, \citenamefont {Cope}, \citenamefont {Spedalieri},\
  and\ \citenamefont {Banchi}}]{pirandola2018theory}%
  \BibitemOpen
  \bibfield  {author} {\bibinfo {author} {\bibfnamefont {S.}~\bibnamefont
  {Pirandola}}, \bibinfo {author} {\bibfnamefont {S.~L.}\ \bibnamefont
  {Braunstein}}, \bibinfo {author} {\bibfnamefont {R.}~\bibnamefont
  {Laurenza}}, \bibinfo {author} {\bibfnamefont {C.}~\bibnamefont {Ottaviani}},
  \bibinfo {author} {\bibfnamefont {T.~P.}\ \bibnamefont {Cope}}, \bibinfo
  {author} {\bibfnamefont {G.}~\bibnamefont {Spedalieri}}, \ and\ \bibinfo
  {author} {\bibfnamefont {L.}~\bibnamefont {Banchi}},\ }\href@noop {}
  {\bibfield  {journal} {\bibinfo  {journal} {Quantum Science and Technology}\
  }\textbf {\bibinfo {volume} {3}},\ \bibinfo {pages} {035009} (\bibinfo {year}
  {2018})}\BibitemShut {NoStop}%
\bibitem [{\citenamefont {Christandl}\ and\ \citenamefont
  {M{\"u}ller-Hermes}(2017)}]{christandl2017relative}%
  \BibitemOpen
  \bibfield  {author} {\bibinfo {author} {\bibfnamefont {M.}~\bibnamefont
  {Christandl}}\ and\ \bibinfo {author} {\bibfnamefont {A.}~\bibnamefont
  {M{\"u}ller-Hermes}},\ }\href@noop {} {\bibfield  {journal} {\bibinfo
  {journal} {Communications in Mathematical Physics}\ }\textbf {\bibinfo
  {volume} {353}},\ \bibinfo {pages} {821} (\bibinfo {year}
  {2017})}\BibitemShut {NoStop}%
\bibitem [{\citenamefont {Brandao}(2008)}]{brandao2008entanglement}%
  \BibitemOpen
  \bibfield  {author} {\bibinfo {author} {\bibfnamefont {F.~G.}\ \bibnamefont
  {Brandao}},\ }\emph {\bibinfo {title} {Entanglement theory and the quantum
  simulation of many-body physics}},\ \href@noop {} {Ph.D. thesis},\ \bibinfo
  {school} {Imperial College London} (\bibinfo {year} {2008})\BibitemShut
  {NoStop}%
\bibitem [{\citenamefont {Yang}\ \emph {et~al.}(2007)\citenamefont {Yang},
  \citenamefont {Horodecki},\ and\ \citenamefont {Wang}}]{yang2007conditional}%
  \BibitemOpen
  \bibfield  {author} {\bibinfo {author} {\bibfnamefont {D.}~\bibnamefont
  {Yang}}, \bibinfo {author} {\bibfnamefont {M.}~\bibnamefont {Horodecki}}, \
  and\ \bibinfo {author} {\bibfnamefont {Z.}~\bibnamefont {Wang}},\ }\href@noop
  {} {\bibfield  {journal} {\bibinfo  {journal} {arXiv preprint
  quant-ph/0701149}\ } (\bibinfo {year} {2007})}\BibitemShut {NoStop}%
\bibitem [{\citenamefont {Yang}\ \emph {et~al.}(2008)\citenamefont {Yang},
  \citenamefont {Horodecki},\ and\ \citenamefont {Wang}}]{yang2008additive}%
  \BibitemOpen
  \bibfield  {author} {\bibinfo {author} {\bibfnamefont {D.}~\bibnamefont
  {Yang}}, \bibinfo {author} {\bibfnamefont {M.}~\bibnamefont {Horodecki}}, \
  and\ \bibinfo {author} {\bibfnamefont {Z.}~\bibnamefont {Wang}},\ }\href@noop
  {} {\bibfield  {journal} {\bibinfo  {journal} {Physical review letters}\
  }\textbf {\bibinfo {volume} {101}},\ \bibinfo {pages} {140501} (\bibinfo
  {year} {2008})}\BibitemShut {NoStop}%
\bibitem [{\citenamefont {Song}(2009)}]{song2009lower}%
  \BibitemOpen
  \bibfield  {author} {\bibinfo {author} {\bibfnamefont {W.}~\bibnamefont
  {Song}},\ }\href@noop {} {\bibfield  {journal} {\bibinfo  {journal}
  {International Journal of Theoretical Physics}\ }\textbf {\bibinfo {volume}
  {48}},\ \bibinfo {pages} {2191} (\bibinfo {year} {2009})}\BibitemShut
  {NoStop}%
\bibitem [{\citenamefont {Huang}(2014)}]{huang2014computing}%
  \BibitemOpen
  \bibfield  {author} {\bibinfo {author} {\bibfnamefont {Y.}~\bibnamefont
  {Huang}},\ }\href@noop {} {\bibfield  {journal} {\bibinfo  {journal} {New
  journal of physics}\ }\textbf {\bibinfo {volume} {16}},\ \bibinfo {pages}
  {033027} (\bibinfo {year} {2014})}\BibitemShut {NoStop}%
\bibitem [{\citenamefont {Wakakuwa}(2019)}]{wakakuwa2019communication}%
  \BibitemOpen
  \bibfield  {author} {\bibinfo {author} {\bibfnamefont {E.}~\bibnamefont
  {Wakakuwa}},\ }\href@noop {} {\bibfield  {journal} {\bibinfo  {journal}
  {arXiv preprint arXiv:1904.08852}\ } (\bibinfo {year} {2019})}\BibitemShut
  {NoStop}%
\bibitem [{\citenamefont {Guha}\ \emph {et~al.}(2007)\citenamefont {Guha},
  \citenamefont {Shapiro},\ and\ \citenamefont {Erkmen}}]{guha2007classical}%
  \BibitemOpen
  \bibfield  {author} {\bibinfo {author} {\bibfnamefont {S.}~\bibnamefont
  {Guha}}, \bibinfo {author} {\bibfnamefont {J.~H.}\ \bibnamefont {Shapiro}}, \
  and\ \bibinfo {author} {\bibfnamefont {B.~I.}\ \bibnamefont {Erkmen}},\
  }\href@noop {} {\bibfield  {journal} {\bibinfo  {journal} {Physical Review
  A}\ }\textbf {\bibinfo {volume} {76}},\ \bibinfo {pages} {032303} (\bibinfo
  {year} {2007})}\BibitemShut {NoStop}%
\bibitem [{\citenamefont {Guha}\ and\ \citenamefont
  {Shapiro}(2007)}]{guha2007classicalproc}%
  \BibitemOpen
  \bibfield  {author} {\bibinfo {author} {\bibfnamefont {S.}~\bibnamefont
  {Guha}}\ and\ \bibinfo {author} {\bibfnamefont {J.~H.}\ \bibnamefont
  {Shapiro}},\ }in\ \href@noop {} {\emph {\bibinfo {booktitle} {Information
  Theory, 2007. ISIT 2007. IEEE International Symposium on}}}\ (\bibinfo
  {organization} {IEEE},\ \bibinfo {year} {2007})\ pp.\ \bibinfo {pages}
  {1896--1900}\BibitemShut {NoStop}%
\bibitem [{\citenamefont {Guha}\ \emph
  {et~al.}(2008{\natexlab{a}})\citenamefont {Guha}, \citenamefont {Erkmen},\
  and\ \citenamefont {Shapiro}}]{guha2008entropy}%
  \BibitemOpen
  \bibfield  {author} {\bibinfo {author} {\bibfnamefont {S.}~\bibnamefont
  {Guha}}, \bibinfo {author} {\bibfnamefont {B.}~\bibnamefont {Erkmen}}, \ and\
  \bibinfo {author} {\bibfnamefont {J.~H.}\ \bibnamefont {Shapiro}},\ }in\
  \href@noop {} {\emph {\bibinfo {booktitle} {Information Theory and
  Applications Workshop, 2008}}}\ (\bibinfo {organization} {IEEE},\ \bibinfo
  {year} {2008})\ pp.\ \bibinfo {pages} {128--130}\BibitemShut {NoStop}%
\bibitem [{\citenamefont {Guha}\ \emph
  {et~al.}(2008{\natexlab{b}})\citenamefont {Guha}, \citenamefont {Shapiro},\
  and\ \citenamefont {Erkmen}}]{guha2008capacity}%
  \BibitemOpen
  \bibfield  {author} {\bibinfo {author} {\bibfnamefont {S.}~\bibnamefont
  {Guha}}, \bibinfo {author} {\bibfnamefont {J.~H.}\ \bibnamefont {Shapiro}}, \
  and\ \bibinfo {author} {\bibfnamefont {B.}~\bibnamefont {Erkmen}},\ }in\
  \href@noop {} {\emph {\bibinfo {booktitle} {Information Theory, 2008. ISIT
  2008. IEEE International Symposium on}}}\ (\bibinfo {organization} {IEEE},\
  \bibinfo {year} {2008})\ pp.\ \bibinfo {pages} {91--95}\BibitemShut {NoStop}%
\bibitem [{\citenamefont {K{\"o}nig}\ and\ \citenamefont
  {Smith}(2014)}]{konig2014entropy}%
  \BibitemOpen
  \bibfield  {author} {\bibinfo {author} {\bibfnamefont {R.}~\bibnamefont
  {K{\"o}nig}}\ and\ \bibinfo {author} {\bibfnamefont {G.}~\bibnamefont
  {Smith}},\ }\href@noop {} {\bibfield  {journal} {\bibinfo  {journal} {IEEE
  Transactions on Information Theory}\ }\textbf {\bibinfo {volume} {60}},\
  \bibinfo {pages} {1536} (\bibinfo {year} {2014})}\BibitemShut {NoStop}%
\bibitem [{\citenamefont {K{\"o}nig}\ and\ \citenamefont
  {Smith}(2016)}]{konig2016corrections}%
  \BibitemOpen
  \bibfield  {author} {\bibinfo {author} {\bibfnamefont {R.}~\bibnamefont
  {K{\"o}nig}}\ and\ \bibinfo {author} {\bibfnamefont {G.}~\bibnamefont
  {Smith}},\ }\href@noop {} {\bibfield  {journal} {\bibinfo  {journal} {IEEE
  Transactions on Information Theory}\ }\textbf {\bibinfo {volume} {62}},\
  \bibinfo {pages} {4358} (\bibinfo {year} {2016})}\BibitemShut {NoStop}%
\bibitem [{\citenamefont {De~Palma}\ \emph {et~al.}(2014)\citenamefont
  {De~Palma}, \citenamefont {Mari},\ and\ \citenamefont
  {Giovannetti}}]{de2014generalization}%
  \BibitemOpen
  \bibfield  {author} {\bibinfo {author} {\bibfnamefont {G.}~\bibnamefont
  {De~Palma}}, \bibinfo {author} {\bibfnamefont {A.}~\bibnamefont {Mari}}, \
  and\ \bibinfo {author} {\bibfnamefont {V.}~\bibnamefont {Giovannetti}},\
  }\href
  {http://www.nature.com/nphoton/journal/v8/n12/full/nphoton.2014.252.html}
  {\bibfield  {journal} {\bibinfo  {journal} {Nature Photonics}\ }\textbf
  {\bibinfo {volume} {8}},\ \bibinfo {pages} {958} (\bibinfo {year}
  {2014})}\BibitemShut {NoStop}%
\bibitem [{\citenamefont {De~Palma}\ \emph {et~al.}(2015)\citenamefont
  {De~Palma}, \citenamefont {Mari}, \citenamefont {Lloyd},\ and\ \citenamefont
  {Giovannetti}}]{de2015multimode}%
  \BibitemOpen
  \bibfield  {author} {\bibinfo {author} {\bibfnamefont {G.}~\bibnamefont
  {De~Palma}}, \bibinfo {author} {\bibfnamefont {A.}~\bibnamefont {Mari}},
  \bibinfo {author} {\bibfnamefont {S.}~\bibnamefont {Lloyd}}, \ and\ \bibinfo
  {author} {\bibfnamefont {V.}~\bibnamefont {Giovannetti}},\ }\href
  {http://journals.aps.org/pra/abstract/10.1103/PhysRevA.91.032320} {\bibfield
  {journal} {\bibinfo  {journal} {Physical Review A}\ }\textbf {\bibinfo
  {volume} {91}},\ \bibinfo {pages} {032320} (\bibinfo {year}
  {2015})}\BibitemShut {NoStop}%
\bibitem [{\citenamefont {De~Palma}\ \emph
  {et~al.}(2016{\natexlab{a}})\citenamefont {De~Palma}, \citenamefont
  {Trevisan},\ and\ \citenamefont {Giovannetti}}]{de2015passive}%
  \BibitemOpen
  \bibfield  {author} {\bibinfo {author} {\bibfnamefont {G.}~\bibnamefont
  {De~Palma}}, \bibinfo {author} {\bibfnamefont {D.}~\bibnamefont {Trevisan}},
  \ and\ \bibinfo {author} {\bibfnamefont {V.}~\bibnamefont {Giovannetti}},\
  }\href {\doibase 10.1109/TIT.2016.2547426} {\bibfield  {journal} {\bibinfo
  {journal} {IEEE Transactions on Information Theory}\ }\textbf {\bibinfo
  {volume} {62}},\ \bibinfo {pages} {2895} (\bibinfo {year}
  {2016}{\natexlab{a}})}\BibitemShut {NoStop}%
\bibitem [{\citenamefont {De~Palma}\ \emph
  {et~al.}(2016{\natexlab{b}})\citenamefont {De~Palma}, \citenamefont {Mari},
  \citenamefont {Lloyd},\ and\ \citenamefont {Giovannetti}}]{de2016passive}%
  \BibitemOpen
  \bibfield  {author} {\bibinfo {author} {\bibfnamefont {G.}~\bibnamefont
  {De~Palma}}, \bibinfo {author} {\bibfnamefont {A.}~\bibnamefont {Mari}},
  \bibinfo {author} {\bibfnamefont {S.}~\bibnamefont {Lloyd}}, \ and\ \bibinfo
  {author} {\bibfnamefont {V.}~\bibnamefont {Giovannetti}},\ }\href
  {http://journals.aps.org/pra/abstract/10.1103/PhysRevA.93.062328} {\bibfield
  {journal} {\bibinfo  {journal} {Physical Review A}\ }\textbf {\bibinfo
  {volume} {93}},\ \bibinfo {pages} {062328} (\bibinfo {year}
  {2016}{\natexlab{b}})}\BibitemShut {NoStop}%
\bibitem [{\citenamefont {De~Palma}\ \emph
  {et~al.}(2017{\natexlab{a}})\citenamefont {De~Palma}, \citenamefont
  {Trevisan},\ and\ \citenamefont {Giovannetti}}]{de2016gaussian}%
  \BibitemOpen
  \bibfield  {author} {\bibinfo {author} {\bibfnamefont {G.}~\bibnamefont
  {De~Palma}}, \bibinfo {author} {\bibfnamefont {D.}~\bibnamefont {Trevisan}},
  \ and\ \bibinfo {author} {\bibfnamefont {V.}~\bibnamefont {Giovannetti}},\
  }\href {http://ieeexplore.ieee.org/document/7707386/} {\bibfield  {journal}
  {\bibinfo  {journal} {IEEE Transactions on Information Theory}\ }\textbf
  {\bibinfo {volume} {63}},\ \bibinfo {pages} {728} (\bibinfo {year}
  {2017}{\natexlab{a}})}\BibitemShut {NoStop}%
\bibitem [{\citenamefont {De~Palma}\ \emph
  {et~al.}(2017{\natexlab{b}})\citenamefont {De~Palma}, \citenamefont
  {Trevisan},\ and\ \citenamefont {Giovannetti}}]{de2016gaussiannew}%
  \BibitemOpen
  \bibfield  {author} {\bibinfo {author} {\bibfnamefont {G.}~\bibnamefont
  {De~Palma}}, \bibinfo {author} {\bibfnamefont {D.}~\bibnamefont {Trevisan}},
  \ and\ \bibinfo {author} {\bibfnamefont {V.}~\bibnamefont {Giovannetti}},\
  }\href {https://journals.aps.org/prl/abstract/10.1103/PhysRevLett.118.160503}
  {\bibfield  {journal} {\bibinfo  {journal} {Physical Review Letters}\
  }\textbf {\bibinfo {volume} {118}},\ \bibinfo {pages} {160503} (\bibinfo
  {year} {2017}{\natexlab{b}})}\BibitemShut {NoStop}%
\bibitem [{\citenamefont {De~Palma}(2016)}]{de2017gaussian}%
  \BibitemOpen
  \bibfield  {author} {\bibinfo {author} {\bibfnamefont {G.}~\bibnamefont
  {De~Palma}},\ }\emph {\bibinfo {title} {Gaussian optimizers and other topics
  in quantum information}},\ \href {https://arxiv.org/abs/1710.09395} {Ph.D.
  thesis},\ \bibinfo  {school} {Scuola Normale Superiore}, \bibinfo {address}
  {Pisa (Italy)} (\bibinfo {year} {2016}),\ \bibinfo {note} {supervisor: Prof.
  Vittorio Giovannetti; arXiv:1710.09395}\BibitemShut {NoStop}%
\bibitem [{\citenamefont {De~Palma}\ \emph
  {et~al.}(2017{\natexlab{c}})\citenamefont {De~Palma}, \citenamefont
  {Trevisan},\ and\ \citenamefont {Giovannetti}}]{de2017multimode}%
  \BibitemOpen
  \bibfield  {author} {\bibinfo {author} {\bibfnamefont {G.}~\bibnamefont
  {De~Palma}}, \bibinfo {author} {\bibfnamefont {D.}~\bibnamefont {Trevisan}},
  \ and\ \bibinfo {author} {\bibfnamefont {V.}~\bibnamefont {Giovannetti}},\
  }\href {https://arxiv.org/abs/1705.00499} {\bibfield  {journal} {\bibinfo
  {journal} {arXiv preprint arXiv:1705.00499}\ } (\bibinfo {year}
  {2017}{\natexlab{c}})}\BibitemShut {NoStop}%
\bibitem [{\citenamefont {Qi}\ \emph {et~al.}(2017)\citenamefont {Qi},
  \citenamefont {Wilde},\ and\ \citenamefont {Guha}}]{qi2017minimum}%
  \BibitemOpen
  \bibfield  {author} {\bibinfo {author} {\bibfnamefont {H.}~\bibnamefont
  {Qi}}, \bibinfo {author} {\bibfnamefont {M.~M.}\ \bibnamefont {Wilde}}, \
  and\ \bibinfo {author} {\bibfnamefont {S.}~\bibnamefont {Guha}},\ }\href@noop
  {} {\bibfield  {journal} {\bibinfo  {journal} {arXiv preprint
  arXiv:1607.05262}\ } (\bibinfo {year} {2017})}\BibitemShut {NoStop}%
\bibitem [{\citenamefont {De~Palma}\ \emph
  {et~al.}(2018{\natexlab{a}})\citenamefont {De~Palma}, \citenamefont
  {Trevisan},\ and\ \citenamefont {Giovannetti}}]{de2018pq}%
  \BibitemOpen
  \bibfield  {author} {\bibinfo {author} {\bibfnamefont {G.}~\bibnamefont
  {De~Palma}}, \bibinfo {author} {\bibfnamefont {D.}~\bibnamefont {Trevisan}},
  \ and\ \bibinfo {author} {\bibfnamefont {V.}~\bibnamefont {Giovannetti}},\
  }\href {\doibase 10.1007/s00023-018-0703-5} {\bibfield  {journal} {\bibinfo
  {journal} {Annales Henri Poincar{\'e}}\ }\textbf {\bibinfo {volume} {19}},\
  \bibinfo {pages} {2919} (\bibinfo {year} {2018}{\natexlab{a}})}\BibitemShut
  {NoStop}%
\bibitem [{\citenamefont {De~Palma}\ \emph
  {et~al.}(2018{\natexlab{b}})\citenamefont {De~Palma}, \citenamefont
  {Trevisan}, \citenamefont {Giovannetti},\ and\ \citenamefont
  {Ambrosio}}]{de2018gaussian}%
  \BibitemOpen
  \bibfield  {author} {\bibinfo {author} {\bibfnamefont {G.}~\bibnamefont
  {De~Palma}}, \bibinfo {author} {\bibfnamefont {D.}~\bibnamefont {Trevisan}},
  \bibinfo {author} {\bibfnamefont {V.}~\bibnamefont {Giovannetti}}, \ and\
  \bibinfo {author} {\bibfnamefont {L.}~\bibnamefont {Ambrosio}},\ }\href
  {https://aip.scitation.org/doi/10.1063/1.5038665} {\bibfield  {journal}
  {\bibinfo  {journal} {Journal of Mathematical Physics}\ }\textbf {\bibinfo
  {volume} {59}},\ \bibinfo {pages} {081101} (\bibinfo {year}
  {2018}{\natexlab{b}})}\BibitemShut {NoStop}%
\bibitem [{\citenamefont {{De Palma}}(2019)}]{de2019new}%
  \BibitemOpen
  \bibfield  {author} {\bibinfo {author} {\bibfnamefont {G.}~\bibnamefont {{De
  Palma}}},\ }\href {\doibase 10.1109/TIT.2019.2914434} {\bibfield  {journal}
  {\bibinfo  {journal} {IEEE Transactions on Information Theory}\ }\textbf
  {\bibinfo {volume} {65}},\ \bibinfo {pages} {5959} (\bibinfo {year}
  {2019})}\BibitemShut {NoStop}%
\bibitem [{\citenamefont {Bombelli}\ \emph {et~al.}(1986)\citenamefont
  {Bombelli}, \citenamefont {Koul}, \citenamefont {Lee},\ and\ \citenamefont
  {Sorkin}}]{bombelli1986quantum}%
  \BibitemOpen
  \bibfield  {author} {\bibinfo {author} {\bibfnamefont {L.}~\bibnamefont
  {Bombelli}}, \bibinfo {author} {\bibfnamefont {R.~K.}\ \bibnamefont {Koul}},
  \bibinfo {author} {\bibfnamefont {J.}~\bibnamefont {Lee}}, \ and\ \bibinfo
  {author} {\bibfnamefont {R.~D.}\ \bibnamefont {Sorkin}},\ }\href@noop {}
  {\bibfield  {journal} {\bibinfo  {journal} {Physical Review D}\ }\textbf
  {\bibinfo {volume} {34}},\ \bibinfo {pages} {373} (\bibinfo {year}
  {1986})}\BibitemShut {NoStop}%
\bibitem [{\citenamefont {Ferraro}\ \emph {et~al.}(2005)\citenamefont
  {Ferraro}, \citenamefont {Olivares},\ and\ \citenamefont
  {Paris}}]{ferraro2005gaussian}%
  \BibitemOpen
  \bibfield  {author} {\bibinfo {author} {\bibfnamefont {A.}~\bibnamefont
  {Ferraro}}, \bibinfo {author} {\bibfnamefont {S.}~\bibnamefont {Olivares}}, \
  and\ \bibinfo {author} {\bibfnamefont {M.}~\bibnamefont {Paris}},\
  }\href@noop {} {\emph {\bibinfo {title} {Gaussian States in Quantum
  Information}}},\ Napoli series on physics and astrophysics\ (\bibinfo
  {publisher} {Bibliopolis},\ \bibinfo {year} {2005})\BibitemShut {NoStop}%
\bibitem [{\citenamefont {Barnett}\ and\ \citenamefont
  {Radmore}(2002)}]{barnett2002methods}%
  \BibitemOpen
  \bibfield  {author} {\bibinfo {author} {\bibfnamefont {S.}~\bibnamefont
  {Barnett}}\ and\ \bibinfo {author} {\bibfnamefont {P.}~\bibnamefont
  {Radmore}},\ }\href@noop {} {\emph {\bibinfo {title} {Methods in Theoretical
  Quantum Optics}}},\ Oxford Series in Optical and Imaging Sciences\ (\bibinfo
  {publisher} {Clarendon Press},\ \bibinfo {year} {2002})\BibitemShut {NoStop}%
\bibitem [{\citenamefont {Wolf}\ \emph {et~al.}(2006)\citenamefont {Wolf},
  \citenamefont {Giedke},\ and\ \citenamefont {Cirac}}]{wolf2006extremality}%
  \BibitemOpen
  \bibfield  {author} {\bibinfo {author} {\bibfnamefont {M.~M.}\ \bibnamefont
  {Wolf}}, \bibinfo {author} {\bibfnamefont {G.}~\bibnamefont {Giedke}}, \ and\
  \bibinfo {author} {\bibfnamefont {J.~I.}\ \bibnamefont {Cirac}},\ }\href@noop
  {} {\bibfield  {journal} {\bibinfo  {journal} {Physical Review Letters}\
  }\textbf {\bibinfo {volume} {96}},\ \bibinfo {pages} {080502} (\bibinfo
  {year} {2006})}\BibitemShut {NoStop}%
\bibitem [{\citenamefont {Giovannetti}\ \emph
  {et~al.}(2015{\natexlab{a}})\citenamefont {Giovannetti}, \citenamefont
  {Holevo},\ and\ \citenamefont
  {Garc{\'\i}a-Patr{\'o}n}}]{giovannetti2015solution}%
  \BibitemOpen
  \bibfield  {author} {\bibinfo {author} {\bibfnamefont {V.}~\bibnamefont
  {Giovannetti}}, \bibinfo {author} {\bibfnamefont {A.}~\bibnamefont {Holevo}},
  \ and\ \bibinfo {author} {\bibfnamefont {R.}~\bibnamefont
  {Garc{\'\i}a-Patr{\'o}n}},\ }\href@noop {} {\bibfield  {journal} {\bibinfo
  {journal} {Communications in Mathematical Physics}\ }\textbf {\bibinfo
  {volume} {334}},\ \bibinfo {pages} {1553} (\bibinfo {year}
  {2015}{\natexlab{a}})}\BibitemShut {NoStop}%
\bibitem [{\citenamefont {Holevo}(2015)}]{holevo2015gaussian}%
  \BibitemOpen
  \bibfield  {author} {\bibinfo {author} {\bibfnamefont {A.~S.}\ \bibnamefont
  {Holevo}},\ }\href@noop {} {\bibfield  {journal} {\bibinfo  {journal}
  {Russian Mathematical Surveys}\ }\textbf {\bibinfo {volume} {70}},\ \bibinfo
  {pages} {331} (\bibinfo {year} {2015})}\BibitemShut {NoStop}%
\bibitem [{\citenamefont {Giovannetti}\ \emph
  {et~al.}(2015{\natexlab{b}})\citenamefont {Giovannetti}, \citenamefont
  {Holevo},\ and\ \citenamefont {Mari}}]{giovannetti2015majorization}%
  \BibitemOpen
  \bibfield  {author} {\bibinfo {author} {\bibfnamefont {V.}~\bibnamefont
  {Giovannetti}}, \bibinfo {author} {\bibfnamefont {A.~S.}\ \bibnamefont
  {Holevo}}, \ and\ \bibinfo {author} {\bibfnamefont {A.}~\bibnamefont
  {Mari}},\ }\href@noop {} {\bibfield  {journal} {\bibinfo  {journal}
  {Theoretical and Mathematical Physics}\ }\textbf {\bibinfo {volume} {182}},\
  \bibinfo {pages} {284} (\bibinfo {year} {2015}{\natexlab{b}})}\BibitemShut
  {NoStop}%
\bibitem [{\citenamefont {Giovannetti}\ \emph {et~al.}(2014)\citenamefont
  {Giovannetti}, \citenamefont {Garc{\'\i}a-Patr{\'o}n}, \citenamefont {Cerf},\
  and\ \citenamefont {Holevo}}]{giovannetti2014ultimate}%
  \BibitemOpen
  \bibfield  {author} {\bibinfo {author} {\bibfnamefont {V.}~\bibnamefont
  {Giovannetti}}, \bibinfo {author} {\bibfnamefont {R.}~\bibnamefont
  {Garc{\'\i}a-Patr{\'o}n}}, \bibinfo {author} {\bibfnamefont {N.}~\bibnamefont
  {Cerf}}, \ and\ \bibinfo {author} {\bibfnamefont {A.}~\bibnamefont
  {Holevo}},\ }\href@noop {} {\bibfield  {journal} {\bibinfo  {journal} {Nature
  Photonics}\ }\textbf {\bibinfo {volume} {8}},\ \bibinfo {pages} {796}
  (\bibinfo {year} {2014})}\BibitemShut {NoStop}%
\end{thebibliography}%

\end{document}